\documentclass[12pt,preprint]{aastex}
\usepackage{amsmath, amsthm, amssymb}

 \newcommand{\xmm}{{\it
    XMM-Newton}}    
 \newcommand{\chandra}{{\it Chandra}}

  \def\ltsima{$\; \buildrel  <  \over \sim
  \;$}    \def\simlt{\lower.5ex\hbox{\ltsima}}    
\def\gtsima{$\;      \buildrel      >      \over      \sim      \;$}
\def\simgt{\lower.5ex\hbox{\gtsima}}      

  \def\ergss{ergs s$^{-1}$\/}

\begin{document} 
   
\title{Suzaku Monitoring of the Seyfert 1 Galaxy NGC5548: Warm
  Absorber Location and its Implication for Cosmic Feedback}
\author{Krongold, Y.\altaffilmark{1}; Elvis, M\altaffilmark{2};
  Andrade-Velazquez, M.\altaffilmark{1}; Nicastro,
  F.\altaffilmark{2,3,4}; Mathur, S.\altaffilmark{5}; Reeves,
  J.N.\altaffilmark{6}, Brickhouse, N.S.\altaffilmark{2}; Binette, L.\altaffilmark{1}; Jimenez-Bailon,
  E.\altaffilmark{1}; Grupe, D.\altaffilmark{7}; Liu,
  Y.\altaffilmark{2}; McHardy, I.M.\altaffilmark{8};  Minezaki,
  T.\altaffilmark{9}; Yoshii, Y.\altaffilmark{9}; and Wilkes,
  B.\altaffilmark{2}}

\altaffiltext{1}{Instituto de Astronomia, Universidad Nacional
  Autonoma de Mexico, Apartado Postal 70-264, 04510 Mexico DF,
  Mexico.}  
\altaffiltext{2}{Harvard-Smithsonian Center for Astrophysics, 60
Garden  Street, Cambridge MA 02138, USA.}  
\altaffiltext{3}{Osservatorio Astronomico di Roma, INAF, Italy.}
\altaffiltext{4}{Foundation for Research and Technology - Hellas
  (FORTH), University of Crete, Greece.}
\altaffiltext{5}{Ohio State University, 140 West 18th Avenue,
  Columbus, OH 43210, USA.} 
\altaffiltext{6}{Astrophysics Group, School of Physical and
  Geographical Sciences, Keele University, Keele, Staffordshire ST5
  8EH, UK}
\altaffiltext{7}{Department of Astronomy and Astrophysics,
  Pennsylvania State University, 525 Davey Lab, University Park, PA
  16802, USA}
\altaffiltext{8}{School of Physics and Astronomy, University of
  Southampton, Southampton SO17 1BJ, UK} 
\altaffiltext{9}{Institute of Astronomy, School of Science,
  University of Tokyo, Mitaka, Tokyo 181-0015, Japan}


\begin{abstract}
We present a two month Suzaku X-ray monitoring of the Seyfert 1 galaxy NGC
5548. The campaign consists of 7 observations (with exposure time
$\sim$ 30 ks each), separated by $\sim$1
week. This paper focus on the XIS data of NGC 5548. We analyze
the response in the opacity of the gas that forms the well known ionized absorber in this source
to ionizing flux variations. Despite variations by a factor $\sim 4$ in
the impinging continuum, the soft X-ray spectra of the source
show little spectral variations, suggesting no response from the ionized
absorber. A detailed time modeling of the spectra
confirms the lack of opacity variations for an absorbing component with
high ionization ($U_X\approx-0.85$), and high outflow velocity
($v_{out}\approx 1040$ km s$^{-1}$), as the
ionization parameter was found to be consistent with a constant value during the
whole campaign. Instead, the models suggest that the ionization
parameter of a low ionization ($U_X\approx-2.8$), low velocity
($v_{out}\approx 590$ km s$^{-1}$) absorbing component might
be changing linearly with the ionizing flux, as expected for gas in
photoionization equilibrium. However, given the lack of spectral
variations among observations, we consider the variations in this component
as tentative. Using the lack of variations, we set an
upper limit of $n_e<$2.0$\times$10$^7$~cm$^{-3}$ for the electron
density of the gas forming the high
ionization, high velocity component. This implies a large distance
from the continuum source ($R~>~0.033$ pc; $R~>~5000~ R_S$). If the
variations in the low ionization, low velocity component are real,
they imply $n_e>$9.8$\times$10$^4$~cm$^{-3}$ and $R<3$
pc. We discuss our results in terms of
two different scenarios: a large scale outflow originating in the
inner parts of the accretion disk,  or a thermally driven wind originating much farther out.
Given the large distance of the wind, the implied mass outflow rate is
also large ($\dot{M}_w > 0.08 \dot{M}_{accr}$) (the mass outflow is
dominated by the high ionization component). The associated total kinetic
energy deployed by the wind in the host galaxy ($>1.2\times 10^{56}$
erg) can be enough to disrupt the interstellar medium, possibly quenching or regulating large
scale star formation. However, the total mass and energy ejected by
the wind may still be lower than the one required for cosmic
feedback, even when extrapolated to quasar luminosities. Such feedback
would require that we are observing the wind before it is fully accelerated.      

\end{abstract}


\section{Introduction \label{par:intro}}

AGN winds, manifested as warm or ionized absorbers (hereafter WA) in
the X-ray and UV bands, are
found in $\sim 50$\% of both quasars (Piconcelli et al. 2005) and Seyfert
galaxies (Crenshaw et al. 2003). Given their high detection rate and
the presence of transverse motion (Mathur et al. 1995;
Crenshaw et al. 2003, Arav 2004), it is
likely that these systems are present in all AGN, but become visible
only when our line of sight crosses the absorbing material. Thus,
it has become clear that understanding these winds is crucial to understand the structure and
geometry of the nuclear region of quasars (Elvis 2000). 
Theoretical studies have shown that AGN winds can have an important
contribution to cosmic feedback if they carry enough mass and
kinetic energy (e.g. Hopkins et al. 2005; Scannapieco \& Oh 2004; King
2003).

Despite their relevance, little is known with confidence about the origin, geometry,
and structure of  AGN winds. However, many physical properties of the
absorbing gas are now relatively well understood, as X-ray WAs can be well-described by a
few absorbing components in pressure balance, with similar
kinematics (e.g. NGC~3783 Krongold et al. 2003, Netzer et al. 2003;
NGC985, Krongold et al. 2005a, 2009; Mrk 279 Fields
et al. 2007, though see Costantini et al. 2007). This has led to the idea that these winds
form a multi-phase medium. The  most important
remaining unknown  of these systems is the location, as their geometry, structure,
and cosmic impact all depend critically on their distance
to the ionizing source.

Determining the location of the winds is not trivial due to the intrinsic
degeneracy of the electron density ($n_e$) and the distance to the
central engine ($R$) in the equation that defines the two observables:
the ionization parameter of the gas ($U_x = Q_x$/[$4\pi R^2 c n_e$]) and the luminosity of
ionizing photons ($Q_x$). One possible solution to this problem comes
from time evolving photoionization, since the ionization equilibrium
time of the gas ($t_{eq}$, i.e. the timescale that the gas requires to
reach equilibrium with the ionizing source) explicitly depends on its
density ($n_e$) (e.g. Nicastro et al. 1999; Krongold et al. 2007;
among others). By monitoring the response of the gas opacity to
ionizing flux variations it is possible to determine $n_e$ and, thus,
the location of the WA winds (see Krongold et al. 2007 for further
details). Another solution to the above degeneracy
relies on measuring line ratios between metastable and stable levels
of ions (for instance Li-like ions), which are also sensitive to the
electron density (e.g. Gabel et al. 2005). Recently, Rozanska et
al. (2008) developed a new technique to estimate the electron density when
the dominant heating mechanism of the absorbing gas is free-free
absorption by soft X-ray accretion disk photons.   

Using these techniques, AGN absorbing winds have been found at
distances that range from few light days (Krongold et al. 2007)
to hundreds of pc away from the central black hole (Crenshaw et al. 2003). Thus, several locations
have been proposed for the origin of the wind including the accretion
disk (Elvis et al. 2000), the ``obscuring torus'' (Krolik \& Kriss
2001), and the narrow emission line region (e.g. Kinkhabwala et al. 2002;
Ogle et al. 2004). Based on the UV spectra of NGC 3783, Gabel et
al. (2005) find a location at $\sim$25 pc. for the highest
velocity component of the absorber in NGC 3783, and concluded that the
gas producing the other two velocity components should lie within that
radius. Using X-ray spectra, Netzer et al. (2003) and Behar
et al. (2003) placed the absorber in this source at distances
larger than 1-3 pc. Krongold et al. (2005b) reported variations in the opacity of the gas
in this source and set an upper limit
of 6 pc. Reeves et al. (2004) found variability of the Fe XXV and XXVI
absorption lines of the high velocity component, 
and located this hot component within 0.1 pc of the central
engine. Kaastra et al. (2004) reported a
possible detection of an O V metastable line in the X-ray spectra of
Mrk 279, placing the gas at distances of lt-weeks to lt-months of the
continuum source. Turner et al. (2008) reported the presence of a partial
covering absorber in NGC 3516, and associated this wind with the
accretion disk. 



Recently, Krongold et al. (2007, hereafter K07) measured short
timescale changes in the opacity of the
absorbing gas forming the ionized wind in the rapidly variable NLSy 1
galaxy NGC4051. During the $\sim 100$ ks \xmm\ observation, the flux level
of NGC 4051 varied by a factor $\sim 10$ allowing for a robust
determination of the gas response timescale, and thus its electron density
and distance. They found that the gas of the two absorbing components
present in the absorber is close to photoionization equilibrium  on
timescales of a few ks during an ionization episode (i.e. when the
source flux brightens). However, for the high ionization component of
the absorber, during recombination episodes the gas could not reach
photoionization equilibrium on timescales $\sim 20$ ks (it is expected
that $t_{eq}$ during recombination is longer than during ionization
periods). This allowed them to set a distance of 0.5-1.0 lt-days for
the high ionization component of the absorber, and an upper limit
of 3.5 lt-days to the distance of the low ionization one. Given that
the location of the absorber was inconsistent with the location of the
obscuring torus in this source (located at distances $>>$ 12 lt-days,
see K07 for details),
and that the absorber was detected on an accretion disk scale
(distances of a few thousand gravitational radii for the $\sim
1.9\times 10^6$ M$_\odot$ central black
hole in NGC 4051, Peterson et al. 2004), they concluded that the origin of the wind should
be the accretion disk. With the location at hand, K07 suggested a
biconical (funnel shaped) wind possibly connected to the high
ionization broad emission line
region (i.e. consistent with the structure for quasars suggested by
Elvis 2000). The mass outflow rate and kinetic energy of the wind were
found to be too weak to have any influence on their large scale environment. However
extrapolating the values found for NGC 4051 (on the very low end of black
hole mass and luminosity) to powerful quasars, they suggested that
cosmic feedback was possible for such extreme objects.

Clearly, in order to understand AGN and quasar winds, as well as their connection
to other components of the nuclear environment and their possible
cosmic feedback, further location measurements are required, on as many objects as possible. 
Here we present a Suzaku X-ray monitoring campaign for the Seyfert 1 galaxy NGC 5548,
designed to follow WA changes in response to continuum variations. The
paper is organized as follows: In \S \ref{par:data}
we describe the reduction of the data. In \S \ref{search}, we
present a search for variability in the spectra of the Suzaku data of NGC
5548. In \S \ref{sec:analysis} we present the data analysis.
In \S \ref{sec:follow} the response of the gas to the continuum
variations is further tested. In \S \ref{sec:absorber} we present the physical conditions and
constrain  the location of the absorber, and in \S \ref{feedback} we discuss its possible
cosmic implications. Finally, in \S \ref{sec:conc} we present our conclusions.

\subsection{The Ionized Absorber in NGC 5548}

NGC5548 ($cz=5149\pm{7}$ km s$^{-1}$; De Vaucouleurs,  1991, based on HI measurements) has a black
hole mass of $6.7\times10^7$M$_\odot$ (Peterson et al. 2004) and an
X-ray 
luminosity of L$_{2-6keV}~=~(2.0~-~2.8)\times~10^{43}$~erg~s$^{-1}$
(Andrade Velazquez et al. 2009, hereafter
AV09). NGC 5548 has a well
known warm absorber. This WA is composed of two velocity components, each with 
two absorbing gas phases (AV09). It has been observed for $\sim 940$ ks
with the \chandra\ gratings.

AV09 modeled the first 800 ks \chandra\ data of NGC 5548,
consisting of $\sim 240$ ks obtained with the HETG and $\sim$ 560 ks
with the LETG (see their Table 1)\footnote{The remaining $\sim$137 ks
  (not modelled by AV09) were obtained $\sim$10
days apart from our Suzaku campaign, but the data were acquired during a
``low flux state'' of NGC 5548 resulting  in spectra of very poor S/N.}. AV09 resolved two velocity systems
in the \chandra\ spectra (as did Steenbrugge et
al. 2005) which turn out to correspond to two out of the five CIV absorbing systems observed in the
UV (Crenshaw et al. 2009). Each X-ray velocity system consisted of two absorbing components. The high
velocity (HV) one (-1040 km s$^{-1}$) consists of a super high ionization component
(or phase, hereafter HV-SHIP), with temperature of $(3\pm0.7)\times10^6$ K
($\log U_x =-1.02\pm0.05$) and a high ionization phase (hereafter HV-HIP) with temperature
of $(7.8\pm1.1)\times10^5$ K ($\log U_x =-1.58\pm0.09$). The
high-velocity, super high-ionization
component produces absorption from charge states
FeXXIII-XXIV, NeX, and, OVIII while the high-velocity, high-ionization component
produces absorption by   MgXI, and OVIII.
The low-velocity (LV) system (-570 km s$^{-1}$) has one high ionization phase (LV-HIP) 
with temperature of $(7.8\pm1.1)\times10^5$ K ($\log U_x =-1.58\pm0.02$)), producing absorption by NeX,
MgXII, OVIII; and a low ionization phase (LV-LIP) with
temperature of $(3.5\pm0.15)\times10^4$ K ($\log U_x =-2.74\pm0.09$) and
producing absorption by OVII, OVI, and the Fe M-shell
UTA (AV09). 


The warm absorber in this source is one
of the few where a model consisting of a few components in pressure
balance was originally disfavored (Steenbrugge et al. 2005),
probably because the two velocity systems were fit together. However, AV09 has
shown that, when the absorbers of each velocity component are modeled
independently, the WA absorber can be well described by two velocity
systems, each consisting of a two phase medium in pressure balance  
(for a detailed analysis of the WA in NGC 5548 see AV09). A few
component, pressure balance model, thus seems generally applicable to WAs.



\section{Observations and Data Reduction  \label{par:data}}

NGC 5548 was observed by Suzaku (Mitsuda et al. 2007), as part of a
$\sim$ 2 month monitoring campaign. In this work we focus on the XIS
data of NGC 5548. The study of the Fe K$_\alpha$ line will be presented in a forthcoming paper
(Liu et al. 2010). Simultaneous Swift observations will
be presented by Grupe et al. (in preparation).

Suzaku observed NGC 5548 seven times
between 18 June 2007 and 6 August 2007 at approximately weekly
intervals (with the exception of the week of 30 June), with net
exposures of typically 25-35 ks per observation. Table \ref{table:log}
presents the log of observations, exposure time, count rate, and S/N
ratio of the data. All observations were performed in the
on-axis XIS nominal pointing position. Data from the X-ray
Imaging Spectrometer ({\sc xis}; Koyama et al. 2007)  were analyzed using
revision 2 of the Suzaku pipeline\footnote{http://www.astro.isas.ac.jp/suzaku/analysis/}. Data were excluded within
436 seconds of passage through the
South Atlantic Anomaly (SAA) and within Earth elevation angles and Bright
Earth angles of $<5^\circ$ and $<20^\circ$, respectively.
A cut-off rigidity (COR) of 6\,GeV was applied, in order to lower the
particle background.

XIS data were selected in $3 \times 3$ and $5 \times 5$
editmodes using grades 0,2,3,4,6, while hot and flickering pixels were
removed using the {\sc sisclean} script. Source spectra were extracted
from within circular regions of 2.9\arcmin\ diameter, while background
spectra were extracted from circles offset from the source and
avoiding the chip corners that contain the calibration sources. The
response matrix ({\sc rmf}) and ancillary response ({\sc arf}) files
were created using the tasks {\sc xisrmfgen} and {\sc xissimarfgen}, respectively,
the former accounting for the CCD charge injection and the latter
for the hydrocarbon contamination on the optical blocking filter.
Spectra from the two front illuminated XIS\,0 and XIS\,3 chips were
combined to create a single source spectrum (hereafter XIS--FI), while
data from the back illuminated  XIS\,1 chip were analyzed separately. Data were included from
0.45--10\,keV for the XIS--FI and 0.45--7\,keV for the XIS\,1 chip,
the latter being optimized for the soft X-ray band. The nominal resolution of both detectors is $\sim 75 eV$ at 2 keV. However,
the XIS spectra were subsequently binned to a minimum of 25 counts per
bin. $\chi^{2}$ minimization was used for all subsequent spectral fitting.
Net count rates for the  XIS--FI range from 0.81 counts\,s$^{-1}$ (lowest) to 3.60 counts\,s$^{-1}$ (highest), while the background rate was $<4$\%
of the source rate during all the observations.



\section{Monitoring the Gas Response to Continuum
  Variations I: A Search for Spectral Changes \label{search}}

\begin{figure}
\plotone{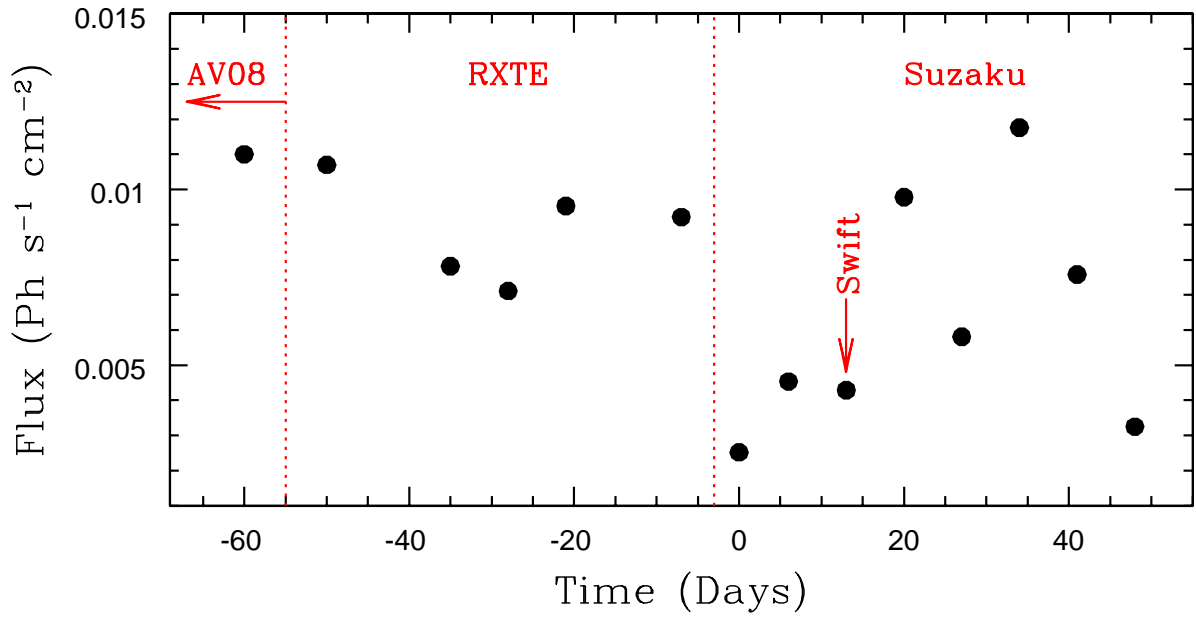} 
\caption[]{0.5 -- 10.0 keV light curve of NGC5548. RXTE and Swift data
  were taken from Liu et al. (2010) and Grupe et al. (2010).
\label{figure:lc}}

\end{figure}

Figure \ref{figure:lc} presents the light curve of the seven Suzaku
observations of NGC 5548. The source varied by a factor of $\sim$4 during
the 2 month monitoring, and up to a factor of $\sim$2 between the 1 week
observations. It is interesting to note also that significant
variability is observed within each observation block. Variations up to 30\%
are detected on typical timescales of $\sim 10$ hours (the typical
exposure times of the observations, see Liu et al. 2010).

\begin{figure}
\plotone{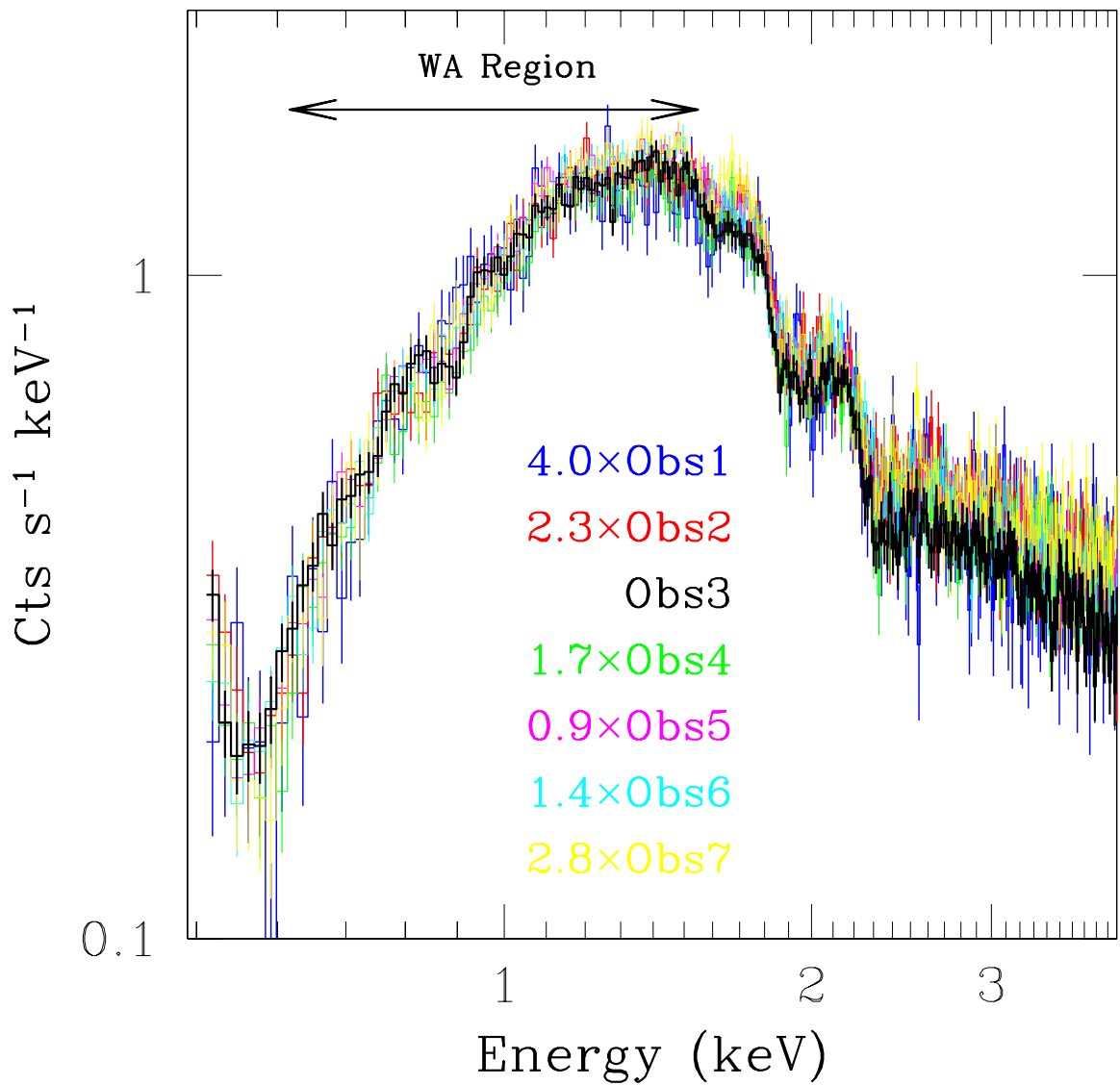} 
\caption[]{XIS-FI Suzaku Spectra of NGC5548. All the spectra are re-scaled
  with respect to Observation 3, for easy comparison. The data are presented
  with nominal binning (though the spectral analysis was carried out
  on data binned including at least 25 cts. per channel). No
  spectral variations are present between 0.6 and 1.6 keV (8 to 20 \AA), indicating no response
  from the WA to the continuum changes. The changes above 2 keV are
  due to variations in the X-ray spectral energy distribution of the
  source (Grupe et al. in preparation).
\label{figure:su_spec}}
\end{figure}

\begin{figure}
\plotone{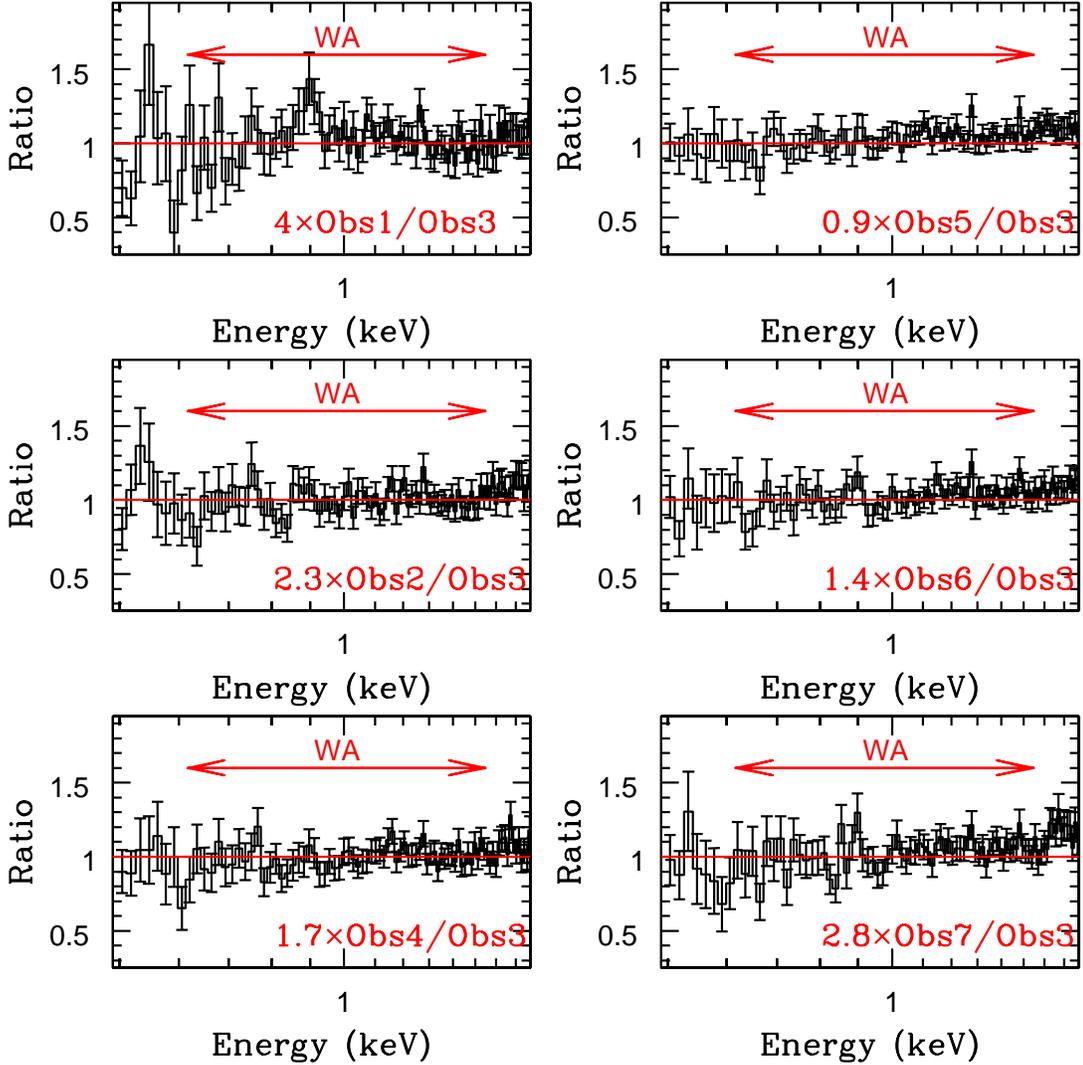} 
\caption[]{Ratio between each XIS-FI observation, and observation 3. The data are presented
  unbinned (though the spectral analysis was carried out on binned
  data). In the Warm Absorber region, between  0.6 and 1.6 keV (8 to 20 \AA), the
  ratios are almost flat, confirming the lack of variability in the opacity of
  the absorbing gas. 
Changes below 0.6 keV  are likely
  produced by the O VII emission triplet (see text for details).  
\label{figure:ratios}}
\end{figure}

In order to look for any possible variations of the ionized absorber to
the variations in the impinging radiation, we have overplotted all the Suzaku spectra
on the same scale. 
This approach is extremely useful when trying to find spectral variations driven by opacity
variations in the absorbing gas, given that re-scaling the spectra allows
to compare directly the transmission of the absorbing gas from
different epochs. We note, however, that the scaling is inappropriate
for comparing emission lines since the flux in these features needs
to have the continuum subtracted from it.  
We have chosen Observation 3 (Table \ref{table:log}) as the
``standard'' normalization point for all observations.
Figure \ref{figure:su_spec} presents the normalized spectra for the
seven Suzaku observations, along with the normalization factors
used to re-scale the spectra. Any variations in the opacity of the gas should be reflected in
the soft X-ray band, especially in the range between 0.6 and 1.6 keV (8 and 20 \AA),
where the majority of the bound-bound and bound-free transitions from
the most abundant metals are located. A striking
result is that all the seven spectra are extremely similar in this
region, indicating that only small spectral variations are
present. This lack of variation is
further evidenced in Figure \ref{figure:ratios}, where the ratios
between the normalized spectra and the spectrum of Observation 3 are
presented. In the soft band, the only regions where some variations are observed
in all spectra are between 0.5 and 0.6 keV (20 and 24 \AA). However, in this region there are not many
absorption lines (the strongest are O VII 21.6\AA, and O VI
22.03\AA). Instead,  it is likely that such variations are
artificially produced by the re-scaling of the spectra, as  the O
VII emission triplet lies in this region. Some other small variations are
present in the soft region on single observations. For instance Observation 6
presents marginal variability near 0.7 and 0.4 keV. Even if such
variations are real, they are too small to consider that the gas can
reach photoionization equilibrium with the ionizing continuum (as we
show below). In the hard region of the
spectra, systematic variations are observed with respect to
Observation 3, indicating a correlation between the flux level and
the X-ray spectral energy distribution of the source (Nicastro et
al. 2000; Grupe et al. in preparation).

\begin{figure}
\plotone{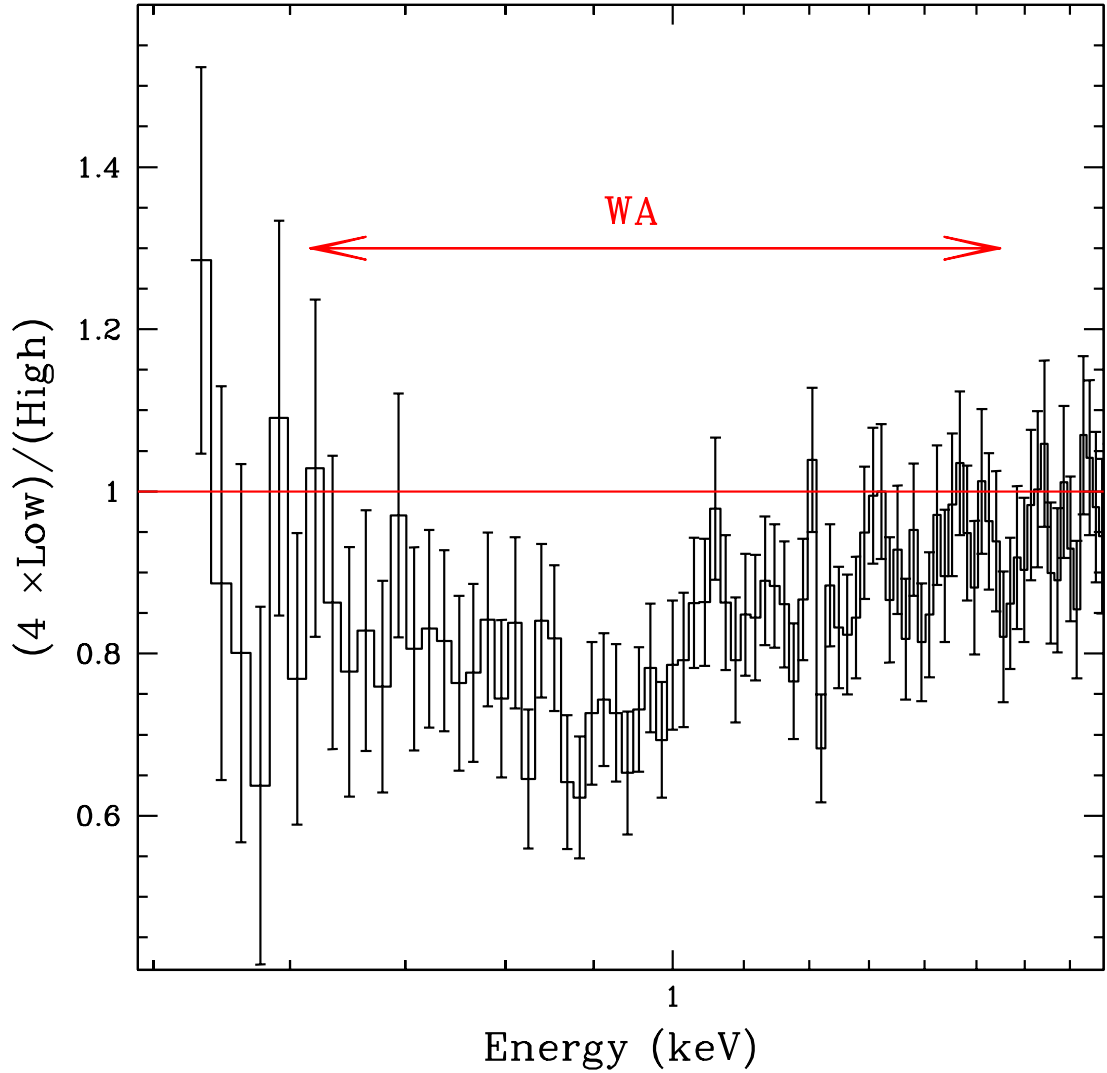} 
\caption[]{Simulated Spectra for the expected spectral variation of an
  absorber in photoionization equilibrium for a factor $\sim$ 4 in ionizing
  flux. The exposure times and continuum levels correspond to Suzaku observations 1
  and 3. Clearly, such changes would be easily detectable in the data.
\label{figure:sim_phase}}
\end{figure}

In Figure \ref{figure:sim_phase} the expected spectral variability for an 
absorber in photoionization equilibrium is presented for a change by a
factor $\sim$4 in the continuum. To produce this plot, we have used
our best-fit model to the data (presented in \S \ref{sec:analysis}),
with the parameters reported for the Suzaku data in Table \ref{table:absorbers}. 
From this plot it is clear that, even if small spectral changes are
present, the gas does not seem to be responding to the
flux changes of the source, and thus, it is likely to be out of
photoionization equilibrium.

\section{Spectral Modeling of the NGC 5548 Suzaku Data. \label{sec:analysis}}

To further test the state of the absorbing gas, and quantify its
physical properties, we have modeled the Suzaku spectra of NGC 5548.
The analysis was carried out with the Sherpa
(Freeman et al. 2001) package\footnote{http://cxc.harvard.edu/sherpa/} of the CIAO software (Fruscione
2002).  The code PHASE (Krongold et al. 2003) has been used to model the ionized absorbers
present in NGC\ 5548. We have explored only photoionization equilibrium
models, adopting the same SED used by AV09 (their Figure 4, mostly obtained from the
NASA Extragalactic Database, NED)\footnote{During the SUZAKU
  monitoring NGC 5548 presented strong changes in the SED (see Grupe
  et al. in preparation). However, the lack of variability in
  the absorber to the X-ray continuum changes makes it reasonable to
  consider a constant SED for the analysis.} . In this paper, we use
the ionization parameter defined in the range 0.1-10 keV ($U_x$, Netzer 1996). this is
different from AV09 who uses the ionization parameter defined over the
whole Lyman range. Given the SED, this two quantities are easily related: $\log U_x
= \log U - 2.25$. 

We have assumed solar elemental abundances (Grevesse et
al. 1993). In all the models, we have taken into account Galactic
absorption by applying a cold absorption component with an
equivalent   Hydrogen column    frozen  to  the   Galactic   value
(N$_H$=1.6$\times10^{20}$\ cm$^{-2}$, Murphy et al. 1996).

Since our goal is to follow the response of the ionized gas present in
the spectra of NGC 5548 using low resolution CCD data, we modeled the
data in a fashion similar to Krongold et al. (2007). These authors
found that when the kinematic properties of ionized absorbers
are constrained by high resolution data, the other properties
(ionization parameter and column density)  can be well
modeled at lower CCD X-ray spectral resolution. Thus, we have modeled the Suzaku XIS spectra of NGC 5548, using the
kinematic properties of the absorbers found by AV09 in  their spectral analysis on the high resolution, high S/N, \chandra\ data.

AV09 found four absorbing components. Two of these show very different
ionization parameters, column densities, and outflow velocities, the HV-SHIP and the
LV-LIP. The other two have 
similar ionization parameters and column densities, but different outflow velocities, the
HV-HIP (with velocity similar to the HV-SHIP) and the LV-HIP (with
velocity similar to the LV-LIP). The low resolution of the Suzaku data
does not allow us to
separate these last two components, even if present in the spectra. Thus,
to test their presence, we model them together, using a
single absorber (we will call this component [HV+LV]-HIP).
In our models of the low resolution CCD Suzaku spectra, we have fixed the outflow
velocity, and turbulent velocity of the HV-SHIP and LV-LIP to the best fit
values obtained by AV09. To account for the
absorption by the [HV+LV]-HIP we have used the average outflow
velocity of the HV-HIP and  LV-LIP systems found by AV09, and a large turbulent velocity (600 km s$^{-1}$, similar to the separation
of the LV and HV systems). The values used for each component are
reported in Table \ref{table:absorbers}. 

Using these three absorbing components
(HV-SHIP, [HV+LV]-HIP, and LV-LIP), we have searched for possible variations in
the opacity of the absorbing gas to the impinging continuum variations. We have used the
following approach: (a) we have assumed that no changes are present
in column densities during the 2 month monitoring campaign. Therefore, we have set free to
vary the column density of each absorbing component, but have forced it
to have the same best fit value during all 7 observations. (b) we have
left the ionization parameter of each absorbing component free to vary
among the observations (as opacity variations should be reflected as variations in the ionization
parameter).

\begin{figure}
\plotone{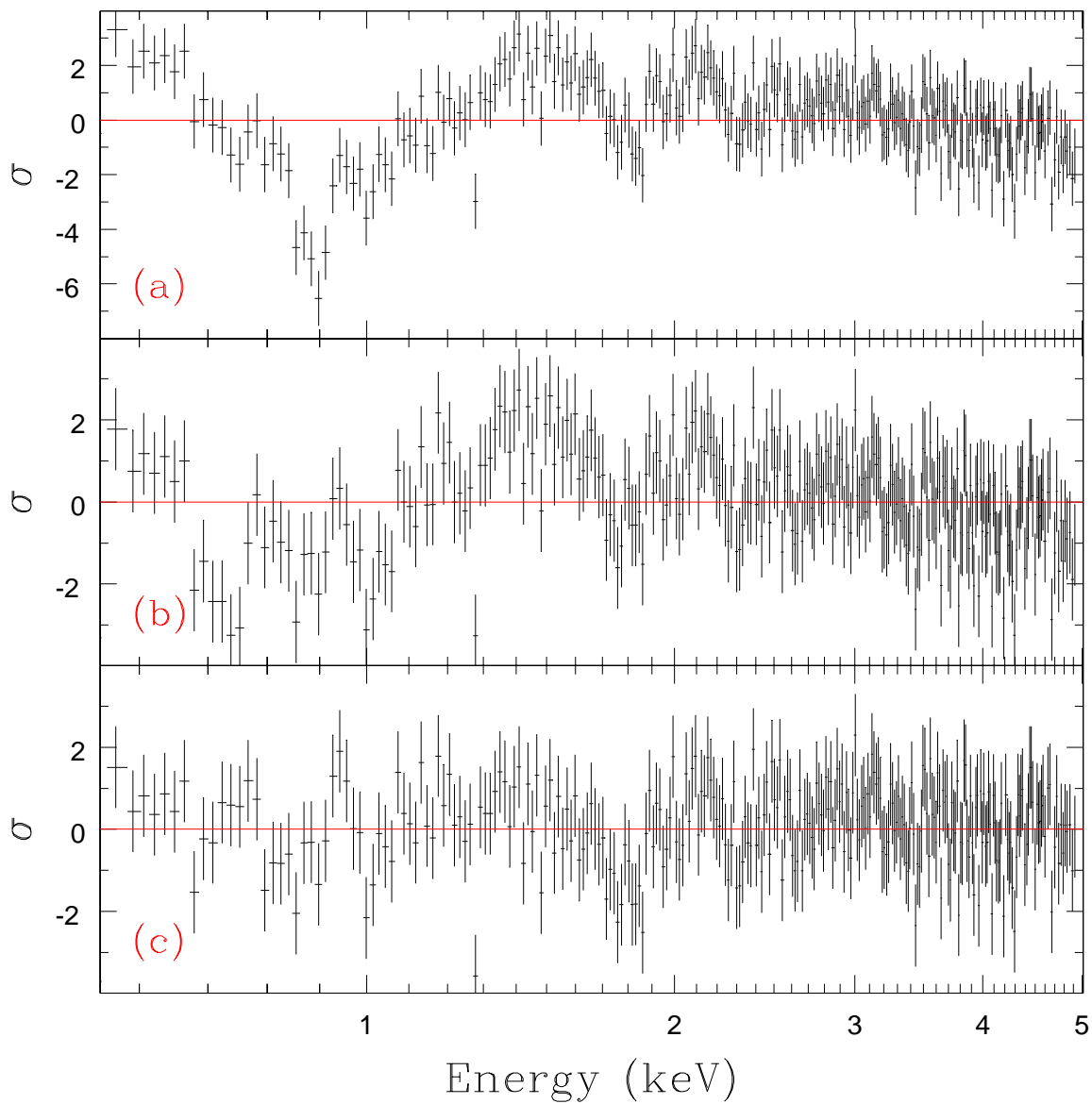} 
\caption[]{Residuals of three different models over Suzaku Obs 3. Panel (a).
  Residuals to a single powerlaw fit between 2 and 5 keV. The presence of the
  ionized absorber is evident in the
  data. Panel (b). Residuals to a fit over the whole spectral range, including a
  single powerlaw attenuated by one absorbing component.  Panel (c). As for
  panel (b) but including two absorbing components. 
\label{fig:suzaku_res}}
\end{figure}

\begin{figure}
\plotone{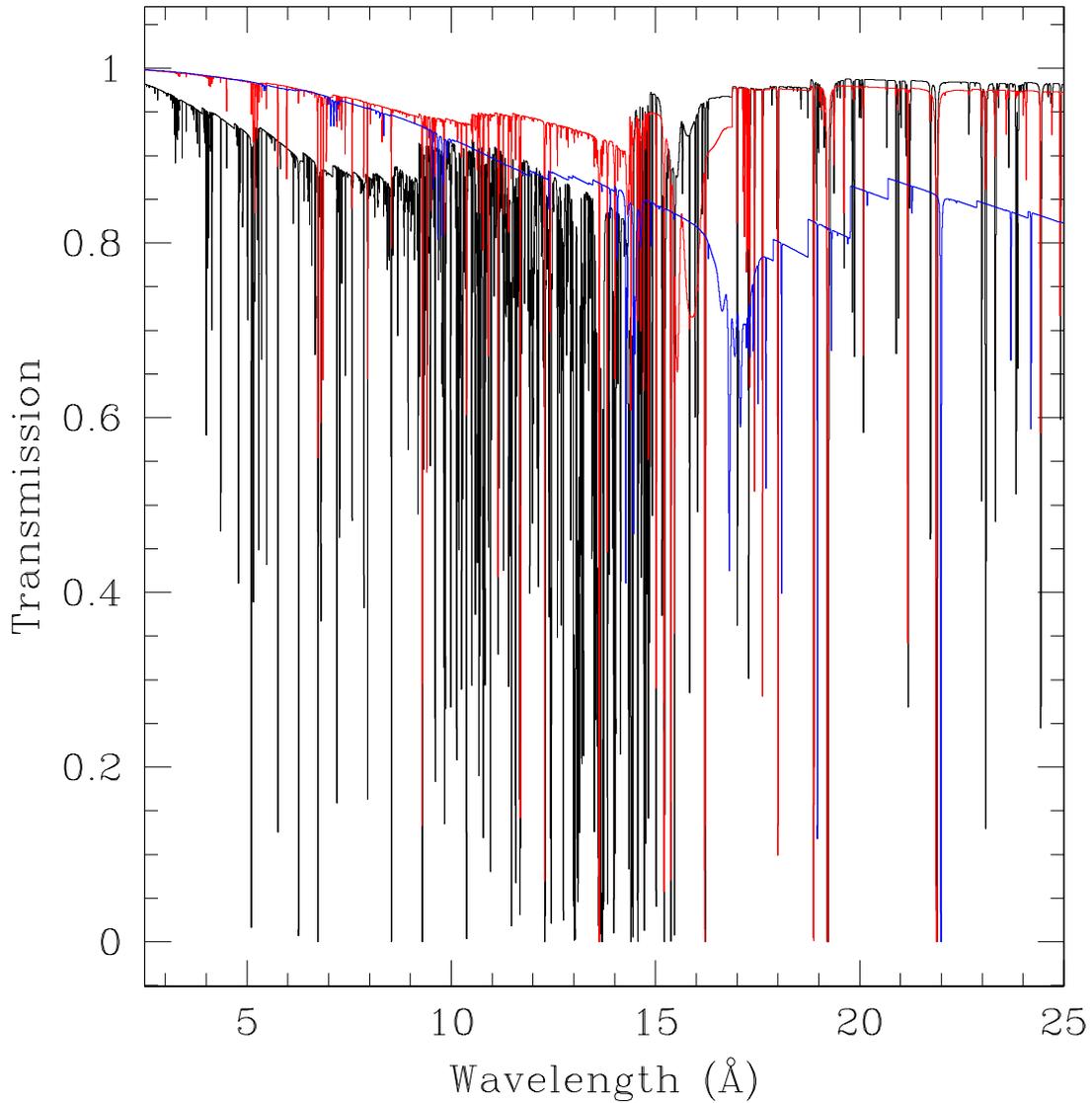} 
\caption[]{Theoretical transmission spectrum
of the three absorption components at 0.001 \AA \ resolution. The HV-SHIP is
presented in black, the [HV+LV]-HIP in red, and the LV-LIP in blue.
While in the 5--15 \AA\ range most of the opacity is produced by the
the HV-SHIP, in the 15--20 \AA\ range the LV-LIP dominates. The
different absorption lines blend into broad troughs that allow us to study
the WA properties with low resolution CCD data (see Fig. \ref{fig:suzaku_res}). 
\label{figure:transmission}}
\end{figure}

We fit simultaneously the XIS-FI and the XIS\,1 spectra of
each observation, and to account for any cross-calibration effect between the
two detectors, we have constrained the slope of the
power law to be the same, but have allowed
the continuum normalizations to vary independently. 
For each observation, we have produced three different models, including a
powerlaw attenuated by one, two, and three absorbing
components (HV-SHIP, [HV+LV]-HIP, and LV-LIP), to test the presence and
significance of each absorber in the Suzaku data. 
The seven datasets require the presence of two absorbing components,
that correspond to the HV-SHIP and the LV-LIP found by AV09. 
Figure \ref{fig:suzaku_res} shows, for Observation 3, the residuals 
to a model including only a power law (Fig. \ref{fig:suzaku_res}a), a powerlaw
attenuated by the HV-SHIP (Fig. \ref{fig:suzaku_res}b), and a powerlaw
attenuated by the HV-SHIP and the LV-LIP (Fig. \ref{fig:suzaku_res}c). 
We note, however, that only for Suzaku Observations 3, 5, and 6 (with
larger S/N) is the
presence of the[HV+LV]-HIP required by the spectra (although,
all datasets are consistent with the presence of this
component). Though statistically required for the above 3 spectra,
this component does
not leave strong (significant) residuals with respect to the model
with only two absorbing components. This is expected, given that the 
[HV+LV]-HIP has an average ionization parameter close to that of
the HV-SHIP, but nearly an order of magnitude lower column density. Thus, the overall resonant opacity of
the absorber between 5-15\AA\ (0.8- 2.5 keV) is mainly driven by the HV-SHIP, and  
the contribution to the absorption by the HV-HIP is difficult to separate
from that component (see  Figure \ref{figure:transmission}), especially with low
resolution CCD data. On the other hand, the opacity in the region
between 15-20\AA\ (0.6-0.8 keV) is dominated by the LV-LIP. This is probably the
reason why the [HV+LV]-HIP component is required only in the 3
observations with larger S/N. 

We conclude that changes in the ionization
parameter of this component (which represents only the ``average'' ionization
parameter between two different velocity components) cannot be constrained
with the present campaign. Therefore, we have included this component in our
final models of the Suzaku data leaving the ionization parameter and
column density fixed to the best fit
values obtained by AV09\footnote{We note, however, that while this
  component is required to fit Observations 3,5, and 6,
letting the parameters of this component vary has no effect on the results
reported in this paper.} (for the value of the column density we have
used the sum of the HV-HIP and LV-HIP columns, for the ionization
parameter, we have used the average; see  Table \ref{table:absorbers}). This component will not be further discussed in
the rest of the paper. 

\begin{figure}
\plotone{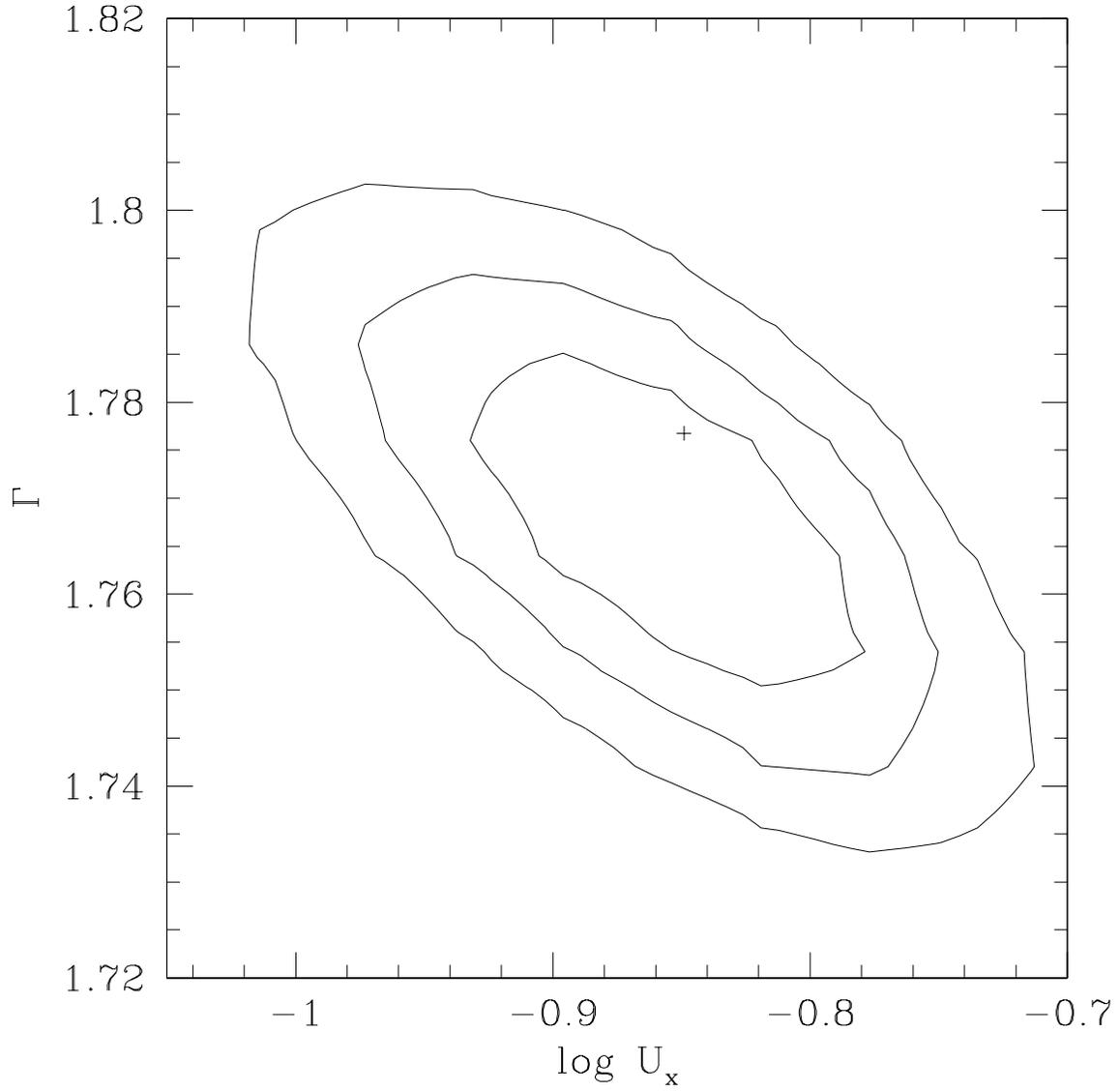} 
\caption[]{Ionization Parameter vs. Photon Index confidence regions
  (1, 2 and 3 $\sigma$) for the fit to Observation 3.  Both parameters
  can be measured independently from the data.
\label{contour}}
\end{figure}

When modeling WAs with low resolution data, an important point to
keep in mind is any possible degeneracy between the continuum and the
opacity of the absorber. Such degeneracy may produce incorrect values
of the ionization parameter because of a bad representation of the
continuum. When monitoring the response of the gas to continuum
variations, such degeneracy may  produce ``artificial changes'' in
$U_x$, or may mask real changes. To test this possible degeneracy we
have produced confidence regions between the photon index of the powerlaw
and $U_x$ (for both the HV-SHIP and the LV-LIP) for the seven Suzaku observations. In Figure \ref{contour}
we present the contour plot for Observation 3 and the HV-SHIP. It is
clear that both the ionization parameter and the  photon index can be
measured independently from the data. Similar results are found for
both the HV-SHIP and the LV-LIP on all observations. A broad band fit
over the XIS and HXD data confirms these results (Liu et al. 2010).    

Table \ref{table:cont} presents the best fit continuum parameters, as
well as the best fit ionization parameters value, of each
observation. Table \ref{table:absorbers} presents the best-fit column
density value of the absorbing components.
We note that the fitted N$_H$ values for these absorbing components
are consistent within the errors with those
found for the \chandra\ best fit model by AV09. This indicates
the reliability of the Suzaku data to constrain this parameter over
the 7 observations. This further indicates no significant change in column density in $\sim$ 7 years of
observations. Using the observed velocities (Table
\ref{table:absorbers}), this means that, if moving purely 
radially, the HV and LV systems have moved at least
$\sim(1-2)\times10^{15}$ cm away from the central source.

\section{Monitoring the Gas Response to Continuum
  Variations II: Time Resolved Modeling of the Ionized Absorber \label{sec:follow}}

\begin{figure}
\plotone{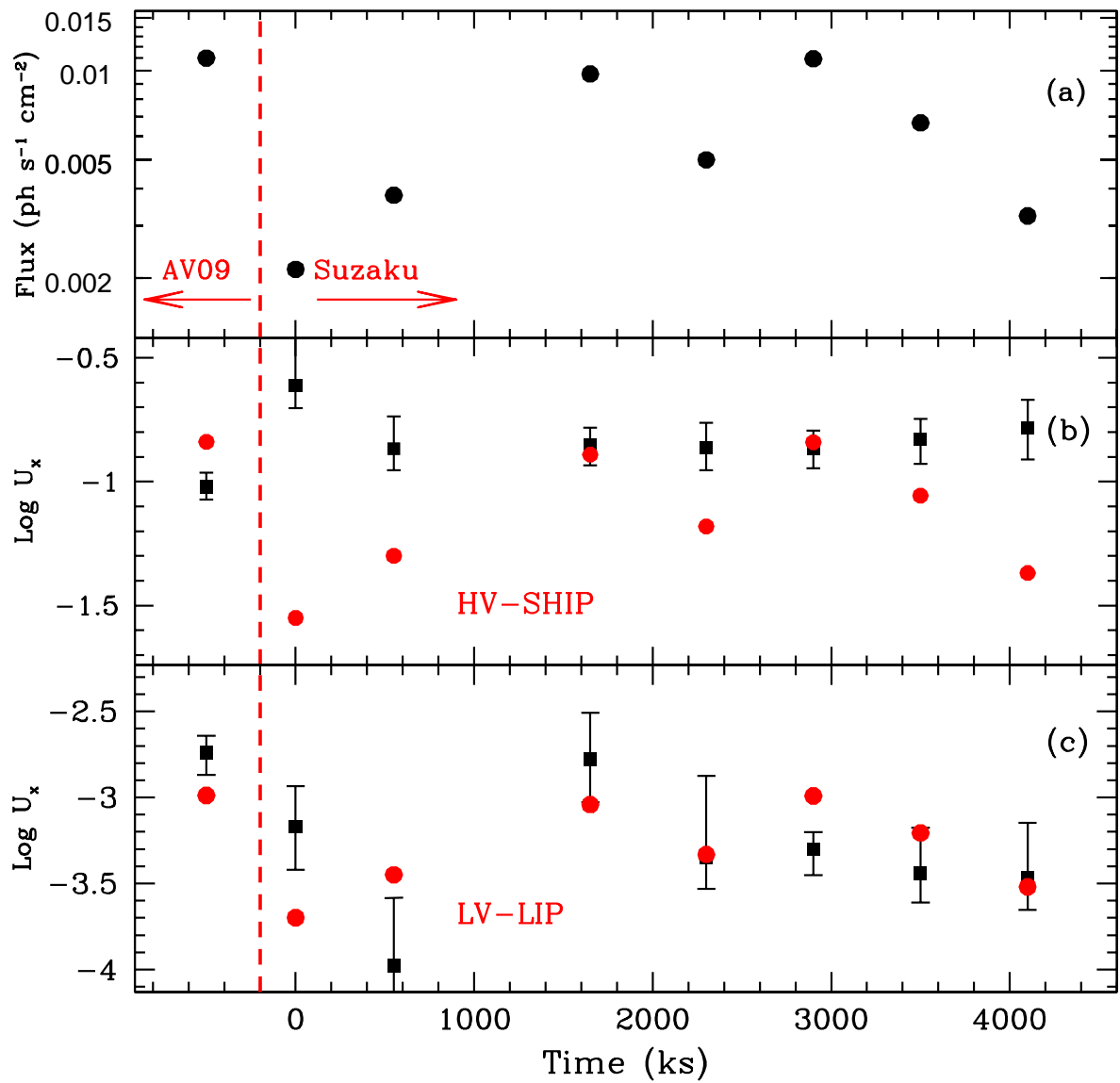} 
\caption[]{Panel (a): Fluxed lightcurve of NGC 5548. The arrows mark the time
  regions when the object was observed by Suzaku and \chandra.  Panel (b):  Log of the Ionization
parameter of the the HV-SHIP vs. time for the 7 Suzaku observations of the
source.
The analysis on the 2000-2005 \chandra\ data
by AV09 is also presented (points with negative time). Panel (c):  as panel (b) but for the
LV-LIP. The red circles represent expectations for gas in photoionization
equilibrium.\label{figure:uvsl}}
\end{figure}

The ionization parameter of the  HV-SHIP and LV-LIP are plotted with the flux level as a
function of time in Figure \ref{figure:uvsl}.
The ionization parameter of the HV-SHIP
is consistent with being constant during the seven Suzaku observations
(Fig. \ref{figure:uvsl}b).
Clearly, the absorbing gas forming
this component is not varying according to the expectations of
photoionization equilibrium (compare Fig \ref{figure:uvsl}a with
\ref{figure:uvsl}b). If the gas were close to photoionization
equilibrium, the  ionization parameter would vary linearly with the flux (as schematically shown by the
red circles in Fig. \ref{figure:uvsl}), which is not the
case. 


Possible changes in $U_x$ are found for the LV-LIP component, which
appears to be following more closely the changes in the continuum
(compare black squares with red circles in  Figure
\ref{figure:uvsl}c). This would indicate that this component is close
to photoionization equilibrium (as all the points are consistent
with the expected values from photoionization equilibrium within
$\sim$2$\sigma$, with the exception of Observation 5, where the difference is
$\sim$2.5$\sigma$). However, those changes are not visible in Figures
\ref{figure:su_spec}  and \ref{figure:ratios}, especially in the
region between 15-20 \AA\ (0.6-0.8 keV), where the Unresolved transition
Array (or UTA) Fe M-shell absorption produced by this component
lies. Furthermore, the errorbars in  $U_x$ for this component are large,
making the values for the seven observations consistent with no variation in
the ionization parameter within 2-3$\sigma$. 
 Thus, it is not clear if
the apparent variations are real, and masked by the non-variability of the
HV-SHIP (which has a column density $\sim 10$ times larger), or if, as for
the [HV+LV]-HIP, the data do not allow us to constrain the response
of the gas. We conclude that time-resolved observations with high
spectral resolution are required to further constrain the behavior of
these two absorbing components. 
Therefore, hereafter we discuss mainly the behavior of the  HV-SHIP,
with a few references to the LV-LIP.

\subsection{Photoionization Equilibrium Timescales \label{sec:time}}

The HV-SHIP did not vary significantly during the two months of
observation, despite the factor $\sim 4$ variations in the flux of the
source. Gas in such conditions, usually remains overionized
with respect to the continuum source during the ``low flux'' states, as $t_{eq}$ during ionization
phases is always shorter than $t_{eq}$ during recombination phases (\S
\ref{par:intro}). This appears to be the case in NGC 5548 where the
HV-SHIP ionization parameter is stuck at high values and does not recombine to reach
photoionization equilibrium during the ``low flux'' observations
(Observations 1,2, 7). This is further supported
by the fact that the ionization parameter of this component is similar
to the one found by AV09 from the average 2000-2005 \chandra\ spectra, with an
``average flux'' (F$_x$[0.5-10 keV]$\sim0.011$ ph s$^{-1}$ cm$^{-2}$) similar to the ``high
flux'' levels seen with Suzaku (Observations 3, and 5, see Fig. \ref{figure:lc})\footnote{We note that the ionization parameter  of the HV-SHIP appears to deviate from photoionization equilibrium by $\sim
  3\sigma$ in the analysis by AV09. However, this effect is likely
  caused by    AV09 modelling the ``average'' 2000-2005 \chandra\ spectra of NGC5548, that
  consists of several datasets with different flux levels (see their Figure
  1).}.  This further supports the idea that the gas is
close to photoionization equilibrium during these ``high
flux'' states, and provides an a-posteriori justification for using
photoionization equilibrium models to fit the spectra. 

Given that we do not have continuous observations of the source, we
cannot determine the exact moment when the continuum decreased to the
``low flux'' states. For these observations (1,2, and 7), the flux
might have decreased just after the observation began, or at any
other time after the previous observation ended. Given that the absorbing gas forming the
HV-SHIP did not responded to continuum variations, we have to assume
the most conservative scenario, which implies that the change in flux
took place just before each ``low state'' observation started. This
means that the gas could not reach photoionization equilibrium with
the continuum in the $\sim$ 30 ks that each of this
observations lasted. Thus we conclude that the  equilibrium time of the
HV-HIP gas is $t_{eq}$(HV-SHIP)$ > 30$ ks. Note that if the flux
decrement took place before the starting point of any of these observations, then
the equilibration time could be $>>$ 30 ks.  


If the LV-LIP is indeed varying as expected in photoionization
equilibrium it can adapt to the continuum changes on a timescale smaller
than the time separation between consecutive observations. Thus $t_{eq}$(LV-LIP)$ < 7$ days ($t_{eq}$(LV-LIP)$ < 0.6\times10^6$ s).

\section{The Large Scale Ionized Absorber in NGC 5548 \label{sec:absorber}}
\subsection{Physical Conditions}

The lower limit on the photoionization timescale of the HV-SHIP
($t_{eq} > 30$ ks) can be
used to set an upper limit on the electron density of the absorbing
gas, as the first is a function of the second (see Nicastro et al 1999
and Krongold et al. 2007 for further details). We have used the approximate relation between
$t_{eq}$ and $n_e$ derived by Nicastro et al. (1999; eq. 5) for a
3-ion atom (i.e. an atom  distributed mainly among three of its contiguous ion species):
\begin{equation*}
t_{eq}^{x{^i},x^{i+1}} \sim
\left[ \frac{1}{\alpha_{rec}(x^i, T_{e})_{eq}~n_e} \right] \times\left[
  \frac{1}{[\alpha_{rec}(x^{i-1}, T_e)/\alpha_{rec}(x^i, T_{e})]_{eq}
    + [n_{x^{i+1}}/n_{x^{i}}]}\right]
\end{equation*} where ``eq'' indicates the equilibrium quantities, and
$\alpha_{rec}(x^i, T_{e}$) is the radiative recombination  coefficient
of the ion $x^{i}$, for gas with an electron temperature $T_e$. This is an 
excellent approximation for O and Ne for the HV-HIP, that has dominant
charge states OVIII-OIX, NeX-NeXI, because 96\% of the population of
these elements is concentrated in these two ions. We used recombination
times from Shull \& van Steenberg (1982) and the average equilibrium
photoionization temperature (see Table \ref{table:phys}). We find
$n_e$(HV-SHIP)$<$2.0$\times$10$^7$~cm$^{-3}$ (we note that the
electron density can be much lower than this value, given the very
conservative limit on the equilibration time). Using $T_e$ and $n_e$, we find that
the gas pressure $P_e$(HV-SHIP)$<5.8\times10^{13}$ K~cm$^{-3}$.  

Assuming the changes in the LV-LIP are real implies
$n_e$(LV-LIP)$>$9.8$\times$10$^4$~cm$^{-3}$ and $P_e$(LV-LIP)$>2.1\times10^{9}$
K~cm$^{-3}$ ($T_e$(LV-LIP)$\sim 2.1\times10^4$ K). 

\subsection{Location and Structure of the Wind \label{sec:loc}}

Using the average ionization parameter of the HV-SHIP during the
observation ($\log U_x=-.83$) and the luminosity of ionizing photons
in the 0.1-10 keV range ($Q_x\sim9.0\times10^{51}$ ph s$^{-1}$), we can
calculate the value of the product ($n_eR^2$): 1.6$\times10^{41}$
cm$^{-1}$ (Table\ref{table:phys}). Given the upper limit on the electron density
found above, the distance of the wind to the central engine can be
derived. We find $R$(HV-SHIP)$>0.033$ pc ($R$(HV-SHIP)$>1.0\times10^{17}$ cm). Given the black
hole mass of NGC 5548 (M$_{BH} = 6.7 \times 10^7$ M$_{\odot}$,
Peterson et al. 2004), this implies a location in Schwarzchild radii $R$(HV-SHIP)$>5.1\times10^3R_s$.   

Assuming a homogeneous flow, and using the column and number
densities it is possible to set constraints on the structure of the ionized
absorber. The line-of-sight thickness can be estimated as $\Delta R \sim N_H /n_H \simeq 1.23 
N_H / n_e$ (where in the last term of the equation we used $n_e \simeq 1.23 
n_H$ which is valid for a fully ionized gas with solar abundances),
and the relative thickness as  $\Delta R / R = 1.23
N_H / n_e R = 1.23 N_H  (n_e R^2)^{-1/2} (n_e)^{-1/2}$. We find 
$\Delta R$(HV-SHIP)$ > 6.1 \times 10^{14}$~cm, and $(\Delta R / R)$(HV-SHIP)$ > 0.007$ (Table
\ref{table:phys}). 

Detmers et al. (2008) found an upper limit of 7 pc for the distance of
the absorber producing the O VIII lines, using long term variations
(in timescales of years) of the source (though the UV absorbers may be
much farther out, i.e. Crenshaw et al. 2009). The HV-SHIP (given its large column
density) has the dominant contribution to the absorption by this
ion. Locating the absorber at this upper limit of 7 pc would produce a
flow with a  $(\Delta R / R) < 1.5$. Thus, for this component $0.033$ pc
$<R$(HV-SHIP)$<7.0$ pc, and  0.007 $<(\Delta R / R)$(HV-SHIP)$ <1.5$.


For the LV-LIP we find  $R$(LV-LIP)$<3.0$ pc ($R$(LV-LIP)$<10^{19}$
cm, $R$(LV-LIP)$<4.5\times10^{5}R_s$), $\Delta R$(LV-LIP)$ < 1.3
\times 10^{16}$~cm, and  $(\Delta R / R)$(LV-LIP)$ <
7.6\times10^{-4}$, if the putative variations in the opacity of this
component are considered as real. Such location is consistent with
the upper limits set for this component by Detmers et al. (2008,
$R$(LV-LIP)$<19.0$ pc) and Crenshaw et al. (2009, $R$(LV-LIP)$<7.0$ pc).

\subsection{Possible Origin and Geometry of the Ionized Outflow}

Our results do not allow us to constrain in detail the geometry
of the wind. However, the small values of $\Delta R / R$ suggest a compact absorber, and thus it is likely that the structure of the absorber is that of a
continuous flow, seen in a transverse direction (Elvis 2000, Arav
2004, K07). The presence of two velocity
absorbing systems, each forming a multi-phase wind (AV09), further suggests a
transverse outflow. This configuration is consistent with a large scale outflow wind, producing both the absorbing gas, and the extended pc-kpc emitting bicones often found in Seyfert 2 galaxies (Kinkhabwala e al 2002, Sako et al. 2000), as has been
explained by K07. 

The location and geometry for the ionized absorber in NGC 5548 can be interpreted in terms
of two different scenarios: an outflow arising from the innermost
regions of the AGN, or an outflow originating much farther out.


\subsubsection{An Inner Accretion Disk Outflow}

In terms of gravitational radii, the HV-SHIP in NGC5548 can be located
at a similar distance (at $> 5000$ R$_s$) than, for instance, the wind in NGC
4051 (at 2000 to 4500 R$_s$) which was established to be connected to
the inner parts of the accretion disk
by Krongold et al. (2007).  However, this component could also be
located much farther out from the center (given that $t_{eq}$ could be
$>> 30$ ks). 

We note that even if this is the case, this does not rule out an
origin in the inner accretion disk. If all warm absorbers originate in the inner accretion disk,
the fact that different locations have been found (not only between the
winds in these two objects, but in many others, see \S \ref{par:intro}) may be
simply reflecting that the wind forms in the accretion disk, but extends over a large scale
in the nuclear (or maybe even galactic) environment of the active galaxy. 
We note that a large scale outflow requires that the wind escapes the gravitational well of the
central black hole. In order to escape, the wind does not
necessarily have to move at escape velocity initially (as stated by several
authors), as long as the radiation force
on the gas from the continuum source is larger than the gravitational
force from the black hole (see Krongold et al. in preparation). Since
both forces depend on the distance as R$^{-2}$, if radiation pressure
dominates at a given distance, it will
dominate at any distance, unless the gas opacity changes dramatically, and the
gravitational force becomes relatively larger. Thus a large scale outflow is
possible even if we do not see the gas moving at escape velocity.

In contrast to NGC 4051, where the absorber's location was consistent with the
Broad Emission Line Region (BELR), in NGC 5548  we are observing the HV-SHIP
farther out. The BELR in this object is located
at least $\sim 2-3$ times closer to the ionizing source (the measured radius is
$\sim 0.01 - 0.02$ pc for
the H$_\beta$ line [Bentz et al. 2006] and 0.014 pc for the He II line
[Bottorff 2002]). This is consistent with UV observations that suggest
that the absorber covers the BELR (Mathur et al. 1995; Crenshaw et
al. 2009). In this scenario, any possible connection between the HV-SHIP and the BELR would
have to be closer to the accretion disk, at the base of the
flow.

The LV-LIP may also be connected to the BELR, since it may be responding in photoionization
equilibrium to the changes in the continuum (see
\S \ref{sec:follow}). In such a case, this
velocity component could be located anywhere within 3 pc from the central
source, and thus, close the base of the flow. Such a location
($\sim0.02$ pc) for this component would imply a photoionization
equilibrium time scale $\sim$ few minutes.  

An intriguing possibility for NGC 5548 is that the two different velocity
components found in absorption correspond to two different locations
where a single disk wind crosses our line of sight to the central source. This is possible for a bi-conical wind,
similar to the one suggested by Elvis (2000) and Krongold et
al. (2007). In this case one of the components could be much farther
out than the other, which should be close to the base of the wind, near
the accretion disk. The relative distance of the two components would
then depend on the bending angle of the funnel-shaped wind with
respect to the accretion disk, and our line of sight angle with
respect to the accretion disk (note that the line of sight angle has to
be larger than the wind angle, in order to cross the flow two times).  
The difference in velocity between the two components does not require strong
acceleration for the flow (though this is possible) given that only
the radial component of the velocity is measured in the
absorbers.

\subsubsection{A Large Scale Origin for the Outflow}

Since ionized absorber winds can arise from both
the accretion disk (Everett 2005; Proga 2004) and from the ``obscuring torus'' (Krolik
\& Kriss 2001, Dorodnitsyn et al 2008), it is possible
that the different locations found for the 
absorbers may be simply reflecting that we are seeing two different outflows
with different origination radius, as suggested by Blustin et
al. (2003). In this case, the different outflow velocity systems found
in NGC 5548 may be due to the different locations. 

\begin{figure}
\plotone{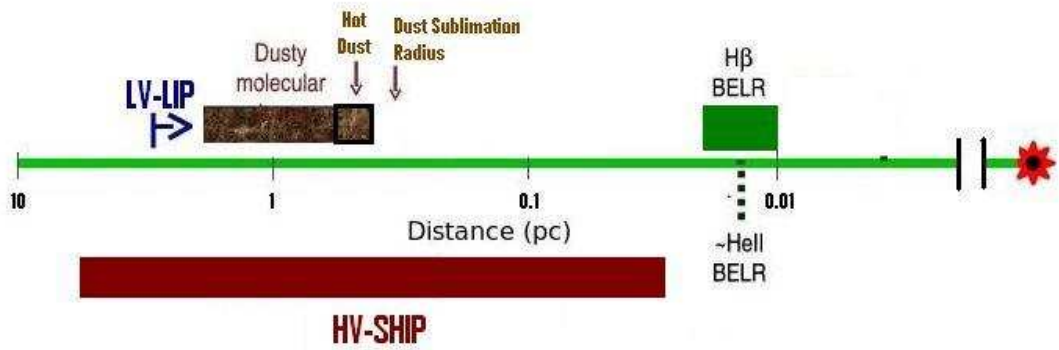} 
\caption[]{Schematic diagram (in logarithmic scale) of the distance to
  the central source of the different AGN components in NGC 5548\label {figure:location}}
\end{figure}

It has been suggested that the ``obscuring torus'' might be part
of an accretion disk wind, whose origin extends from the inner regions
of the accretion disk, up to few pc. In such a scenario, a dusty flow is the  natural continuation
of an inner accretion disk flow beyond the dust sublimation
radius (e.g. Elitzur 2008). Whether the obscuring flow is formed by
clumps (Nenkova et al. 2008) or not (Chang et al. 2007), it must be neutral close to the disk
plane, but could get ionized at larger heights from it. We note
that the lower limit on the location of the HV-SHIP is $\sim 10$ times
smaller than the dusty torus location in NGC 5548
(see Fig. \ref{figure:location}), that has its hotter parts at $\sim$0.4--0.45 pc             from the central source,
(Suganuma et al. 2006), and it is also below the dust sublimation
radius for this object ($\sim$0.35 pc, based on the UV luminosity of NGC 5548 and the formula
given by Barvainis et al. 1987). Thus, the two velocity absorbing
systems in NGC 5548 may arise before the dusty wind (in the inner accretion disk as suggested before).

However, just considering the location constraints found here, the origin of both components could be associated with the ``extended accretion disk dusty flow''. 
If ionized absorbers are indeed connected to this outer dusty accretion disk wind, it would be expected that, at least in some of them, a mix of ionized gas and dust is present (given
that the dust temperature might be much lower than that of the
gas). Indeed, a dusty warm absorber has already been suggested in
MCG-6-30-15 (Lee et al. 2001) and IRAS 13349+2438 (Komossa \&
Breitschwerdt 2000). 

Crenshaw, et al. (2009) have shown that the inner nucleus of NGC~5548 is reddened only slightly: E(B-V) = 0.07, which corresponds to a column density of N$_H\approx3.6\times10^{20}$ cm$^{-2}$  considering a Galactic dust-to-gas ratio. This column density is only a factor $\sim$2 smaller than the column density of the LV-LIP. Thus the LV-LIP may contain some dust. This opens the possibility that this component arises in the ``extended  accretion disk dusty flow'', beyond the dust sublimation radius.  

On the other hand, the column density of the HV-SHIP is much larger than the value inferred from the reddening of the inner nucleus, indicating that this absorbing component cannot contain much dust, and therefore is not a dusty outflow. Therefore, if not located before the dust sublimation radius, the most likely location for the HV-SHIP should be much farther out than the ``extended accretion disk dusty flow'', and then, (if not arising from the inner parts of the disk) this component should be part of a thermally driven wind (Chelouche 2008).

Clearly, these different scenarios could be easily tested with tighter constraints on the
location of the absorbers.

\section{Wind Mass and Kinetic Energy: Implications for Cosmic Feedback \label{feedback}}

The lower limit on the location of the wind makes possible to estimate a
lower limit on the mass and and kinetic energy of the flow.  We can perform
this estimation for a bi-conical wind seen in a
transverse direction with respect to our line of sight. For such a geometry
Krongold et al. (2007) found that 
\begin{equation*}
\dot{M}_w = 0.8 \pi m_p N_H v_r R f(\delta,\phi)
\end{equation*}
where $f(\delta,\phi)$ is a factor that depends on the particular
orientation of the disk and the wind with respect to our line of sight, and for all reasonable angles
is of the order of unity (see Krongold et al. 2007, \S A.2 for full details).
Thus, using the radial velocity, column density, and radius of the HV-SHIP we
find that $\dot{M}_w > 0.008$ M$_\odot$ yr$^{-1}$. The observed accretion 
rate in NGC 5548 is $\dot{M}_{accr} = 0.1$ M$_{\odot}$ yr$^{-1}$
(Mathur et al. 1995, assuming a radiation conversion efficiency of
10\%), so $\dot{M}_w > 0.08 \dot{M}_{accr}$. We note that this is only
a lower limit on the mass outflow rate of the wind, and includes only 1 absorbing
component. If this velocity component is indeed composed of a multi-phase
medium in pressure balance (AV09), the HV-SHIP and HV-HIP should be co-located. In
this case the total mass outflow rate would be  $\dot{M}_w > 0.4
\dot{M}_{accr}$. 
Therefore, the total mass outflow rate of the HV system in NGC 5548 
could easily be much larger than the mass accretion rate. We note that a very large rate of mass outflow could present a
problem for a radiation driven wind, however, the wind may be
magnetocentrifugally launched as suggested by Everett (2005) and
then further accelerated by radiation pressure. 

We stress here that the implication of large mass outflow rates does not depend on the assumed 
bi-conical geometrical configuration. If the outflow were considered spherical,
then the outflow rate, given by $\dot{M}_w=\Omega m_p n_e R^2 V_r$
(e.g. Steenbrugge et al. 2005), would be nearly two orders of magnitude larger,
considering a solid angle of 1.6 ste rad (Blustin et al. 2005). A model
for the WA in this source, assuming a thermally driven wind arising at $\sim$ pc from
the central engine (Chelouche 2008), also implies a much larger mass outflow rate
$\dot{M}_w \sim 0.63$ M$_\odot$ yr$^{-1}$. This mass loss rate is
similar to the one we would obtain for the HV-SHIP, if this component
were located at $\sim$ 1 pc from the central engine ($\dot{M}_w \sim 0.25$ M$_\odot$ yr$^{-1}$).  

Assuming that the black hole mass of NGC 5548 (M$_{BH} = 6.7 \times
10^7$ M$_{\odot}$,  Peterson et al. 2004) was all accreted, and that
the ratio $\dot{M}_w~/~\dot{M}_{accr}$ remained constant, then the
integrated lifetime mass lost due to the HV-SHIP X-ray wind would be
M$_{out} > 5.4 \times 10^6$ $M_{\odot}$. Most of the mass would be
emitted during the last two doubling times, over outburst lifetimes
estimated to be $\sim10^7$ yr. This mass, which comes only
from one absorbing component, is already similar to the one
available from luminous Infrared galaxies (e.g. Veilleux et al. 2005) and,
in principle, could all be injected into the interstellar medium of NGC 5548
or the intergalactic medium. This could have important metalicity effects, as it has
been shown that the active nuclear environment is enriched in metals, with
metalicities a few times solar (e.g. Fields et al. 2005; 2007, Arav et al. 2007) or even larger (Hamman
and Ferland 1999). 

The total kinetic energy injected by the HV-SHIP in NGC 5548 would be $>
1.2\times 10^{56}$ erg. Evaporating the ISM from a typical spiral galaxy with
galactic disk of radius 10 kpc, disk thickness of 0.1 kpc, and average
density of 1 cm$^{-3}$, requires that this medium be heated to a temperature
T$\sim 10^7$ K. The energy needed to increase the ISM of a typical
galaxy to this temperature can be estimated as $E\sim N_{tot}kT \sim 10^{57}$ erg
cm$^{-2}$ s$^{-1}$ (where $ N_{tot}$ is the total number of particles
in the disk). Thus, the kinetic energy injected by the wind can easily
supersede the one required to disrupt the interstellar medium. If AGN activity
occurs intermittently, then the injected kinetic energy would be lower, and the
ISM could have time to fall back. As noted by
Krongold et al. (2007), winds moving at $\sim$500 km s$^{-1}$ can move 5
kpc away in only $10^7$ years (compared to the $10^8$ yr the AGN estimated 
lifetime).  This
implies that AGN warm absorber winds can indeed have important effects on their
host galaxies. Among other things, these winds may control or 
stop large scale star formation processes in their hosts.

If we now assume that the measured ratio $\dot{M}_{out}/\dot{M}_{accr}$ is representative
of quasars, then extrapolating these values for wind mass and kinetic energy
to powerful quasars ($L_{bol} \sim 10^{47}$
\ergss, M$_{BH} = $ few$\times10^9$ M$_{\odot}$), implies that the total mass
injected by these systems in their large scale environment could be
$>$few$\times10^8$ M$_{\odot}$ and the total kinetic energy could be
 $>$few$\times10^{57}$ erg. This number is still a small fraction of the $10^{60}$ erg
 required by simulations to make quasar winds important for galaxy evolution
 and IGM structure formation (e.g. Hopkins et al. 2005; Scannapieco 2004; King
2003). However, the above number is only a lower limit, and includes the
contribution from only one WA component, with a lower value of the
total velocity. In other words, we are taking into account only the radial component of the velocity,
and neglecting further acceleration of the flow and mass
entrainment.

Recent calculations by Hopkins and Elvis
(2009) show that if the AGN wind drives a secondary outflow in the
hot, diffuse gas of the ISM, then a mass outflow rate of only
$\dot{M}_w\sim 0.05-0.10 \dot{M}_{accr}$ for the AGN wind is required
for these outflows to have strong evolutionary effects on their
hosts. Thus it is possible that WA winds may indeed have a
dramatic effect on their large scale environment.  
Evidently, to further understand the cosmological effects of quasar
winds, more theoretical and observational studies are
required. As part of an observational program, we are studying the UV and X-ray
absorbers on the Seyfert galaxy NGC 3227.

\section{Conclusions \label{sec:conc}}

The spectral shape of the soft X-ray spectra of seven Suzaku
observations, spanning two months, shows little variation, despite a
change in flux by a factor $\sim 4$. A detailed
modeling of the ionized absorber in this source confirms that the
strongest component, the one with the largest velocity and
the highest ionization level (HV-SHIP), is not responding as expected in
photoionization equilibrium to the impinging continuum. The lack
of variation in the opacity of this component implies a density
$n_e<$2.0$\times$10$^7$~cm$^{-3}$ for the absorbing gas, and a
distance $>0.033$ pc from the central black
hole. On the other hand, the ionization state of the absorbing component with low velocity and
low ionization (LV-LIP) appears to be varying linearly with the ionizing continuum,
suggesting that it is in photoionization equilibrium. In this case,
this component should have  $n_e>$9.8$\times$10$^4$~cm$^{-3}$ and
should be located anywhere within 3
pc from the central source. 

If the HV-SHIP originates in the inner accretion disk, and we are
seeing it farther out, then a large scale outflow may be
required. Any relation of this component with the BELR would have to
be at the base of the flow, closer in than we are observing
it. If the wind starts perpendicular to the disk, and then bends
(Elvis 2000), the two absorbing components could be part of the
same flow, but at different locations. 
Alternatively, the HV-SHIP could be arising at a much larger radius, and then it might be a thermally driven wind. 

The mass and energy outflow from the wind is dominated by the HV-SHIP. The total kinetic
energy ejected by this wind can easily supersede the energy
required to disrupt the whole interstellar medium of the host
galaxy. This indicates that even winds in Seyferts might have important
effects on their hosts, possibly quenching or regulating large
scale star formation. However, strong uncertainties in the energetics
of these outflows remain. In addition, the total amount of energy required in
these outflows for cosmic feedback could be much lower than
previously thought  (Hopkins and Elvis 2009).  Thus, more studies are
required to further understand the
possible effects of quasar winds on their large scale environment.

\acknowledgements 
This work was supported by the UNAM
PAPIIT grant IN118905 and the CONACyT grant J-49594. NSB acknowledges
support from NASA to the Chandra
X-ray Center through NAS8-03060. This work was supported by
NASA grant NNX08AB81G.   


\clearpage

\clearpage

\begin{deluxetable}{cccccc}
\tablecolumns{6} \tablewidth{0pt}
 \tablecaption{Observation summary of the Suzaku XIS monitoring of NGC
   5548 \label{table:log}} 
\tablehead{\colhead{\bf Obs. Number}&\colhead{\bf Obs. Id.}
  &\colhead{\bf Start Time}  &\colhead{\bf Exp. Time (ks)}
  &\colhead{\bf Ct Rate$^{a}$}  &\colhead{\bf S/N$^{b}$}}
\startdata 
Obs 1 & 702042010 & 2007-06-18 22:28 & 31.1 & 0.81 &  9.4 \\
Obs 2 & 702042020 & 2007-06-24 21:54 & 35.9 & 1.42 & 12.3 \\
Obs 3 & 702042040 & 2007-07-08 10:03 & 30.7 & 2.97 & 18.4 \\
Obs 4 & 702042050 & 2007-07-15 13:58 & 30.0 & 1.83 & 12.6 \\
Obs 5 & 702042060 & 2007-07-22 10:40 & 28.9 & 3.60 & 18.3 \\
Obs 6 & 702042070 & 2007-07-29 04:21 & 31.8 & 2.36 & 15.6 \\
Obs 7 & 702042080 & 2007-08-05 00:38 & 38.8 & 1.23 & 11.4 \\
\enddata
\tablenotetext{a}{Count rate in counts s$^{-1}$ ($0.45~<~E~<~10.0$
  keV). 
The Ct. Rate corresponds to the XIS--FI chip.}
\tablenotetext{b}{Signal to Noise ratio at 15\AA\ (0.83 keV),
given for the spectra with nominal binning obtained of
  the XIS--FI chip. 
}
\end{deluxetable}

\clearpage

\begin{deluxetable}{lccccccccc}
\tablecolumns{10} \tablewidth{0pc} \rotate 
\tabletypesize{\footnotesize}
\tablecaption{Absorber
Parameters \label{table:absorbers}}

\tablehead{\colhead{Observation} & \multicolumn{3}{c}{HV-SHIP$^{c}$}&
  \multicolumn{3}{c}{[HV+LV]-HIP$^{d}$} & \multicolumn{3}{c}{LV-LIP$^{c}$} \\
& \colhead{$\log~U_X$} & \colhead{$\log~N_H$$^{a}$}& \colhead{V$_{out}$$^{b}$}
& \colhead{$\log~U_X$} & \colhead{$\log~N_H$$^{a}$}&
  \colhead{V$_{out}$$^{b}$} & \colhead{$\log~U_X$} & \colhead{$\log~N_H$$^{a}$}& \colhead{V$_{out}$$^{b}$}} 
\startdata
AV09 & $-1.02\pm0.05$ & 21.73$\pm{0.12}$ & 1040$\pm{150}$ &  \nodata & \nodata &
\nodata &  $-2.74\pm0.09$ & 20.75$\pm{0.3}$ & 590$\pm{150}$ \\
HV-HIP (AV09)&\nodata & \nodata & \nodata &  $-1.58\pm0.09$ & 21.03$\pm{0.07}$ &
1180$\pm{150}$ & \nodata & \nodata & \nodata \\ 
LV-HIP (AV09)&\nodata & \nodata & \nodata &  $-1.58\pm0.02$ & 21.26$\pm{0.04}$ &
400$\pm{150}$ & \nodata & \nodata & \nodata \\   
Suzaku$^{e}$ & $-0.85^{+0.07}_{-0.08}$ & 22.0$\pm{0.3}$ & 1040 & -1.58 & 21.46 &
 790  &  $-2.78^{+0.27}_{-0.24}$ &  20.8$\pm{0.3}$ & 590
\enddata
\tablenotetext{a}{In [cm$^{-2}$].} 
\tablenotetext{b}{in km s$^{-1}$. The outflow velocity was constrained
using the best fit values obtained by AV09 over \chandra\ data. The turbulence velocity for the HV-SHIP (160 km
  s$^{-1}$) and LV-LIP (100 km s$^{-1}$) was also constrained using the
  analysis by AV09. The turbulence velocity of the  [HV+LV]-HIP was
  set to 600 km s$^{-1}$, the velocity separation between the HV and
  LV systems.}
\tablenotetext{c}{The column density for the HV-SHIP and LV-LIP was constrained
  to have a single value for all Suzaku observations.
}
\tablenotetext{d}{The column density of the [HV+LV]-HIP from the Suzaku
  data was fixed to the sum of the columns found by AV09 for the
  HV-HIP and LV-HIP.}
\tablenotetext{e}{The values for the ionization parameter reported for the
  Suzaku data correspond to Obs. 3.}

\end{deluxetable}
\clearpage


\begin{deluxetable}{lcccc}
\tablecolumns{5} \tablewidth{0pc} \tablecaption{Continuum, Flux and
  Ionization Parameter for each Suzaku Observation
\label{table:cont}}

\tablehead{\colhead{Observation} & \colhead{$\Gamma$}&
  \colhead{Norm$^{a}$} & \colhead{$log U_x$(HV-SHIP)} & \colhead{$log U_x$(LV-LIP)}} 
\startdata
Obs 1 & 1.51$\pm{0.02}$ &  1.39$\pm{0.05}$ &  
-0.61$^{+0.27}_{-0.09}$ &  -3.17$^{+0.24}_{-0.25}$  \\  
Obs 2 & 1.59$\pm{0.04}$ &  2.80$\pm{0.08}$ &  
-0.87$^{+0.13}_{-0.09}$ &  -3.98$^{+0.39}_{-0.49}$ \\
Obs 3 & 1.78$\pm{0.02}$ &  7.96$\pm{0.04}$ &  
-0.85$^{+0.07}_{-0.08}$ &  -2.78$^{+0.27}_{-0.25}$ \\
Obs 4 & 1.55$\pm{0.01}$ &  3.46$\pm{0.03}$ &  
-0.86$^{+0.10}_{-0.09}$ &  -3.35$^{+0.48}_{-0.18}$ \\
Obs 5 & 1.69$\pm{0.01}$ &  7.95$\pm{0.04}$ & 
-0.86$^{+0.07}_{-0.08}$ &  -3.30$^{+0.10}_{-0.15}$ \\
Obs 6 & 1.59$\pm{0.01}$ &  4.66$\pm{0.03}$ &  
-0.83$^{+0.08}_{-0.10}$ &  -3.44$^{+0.26}_{-0.17}$\\
Obs 7 & 1.56$\pm{0.01}$ &  2.33$\pm{0.02}$ &  
-0.78$^{+0.11}_{-0.13}$ &  -3.47$^{+0.32}_{-0.19}$\\
\enddata
\tablenotetext{a}{At 1 keV, in units of $10^{-3}$ ph cm$^{-2}$
  s$^{-1}$ keV$^{-1}$ }
\end{deluxetable}


\begin{table}
\caption{Physical parameters for the HV-SHIP\label{table:phys}}
\footnotesize
\begin{tabular}{cccccccc}
\hline
N$_H$ & T$_e$ & $(n_eR^2)$ & $n_e$ & P$_e$ & $R$ & 
$\Delta R$ & $(\Delta R / R)$ \\
$10^{22}$ cm$^{-2}$ & $10^6$ K & $10^{41}$ cm$^{-1}$ & $10^{7}$ cm$^{-3}$ &
$10^{13}$ K cm$^{-3}$ 
& $10^{17}$ cm & $10^{14}$ cm  & \\
\hline
$1.0 \pm{2.2}$ & $2.9 \pm{1.4}$ & $1.6\pm 0.05$ & $<$2.0 & $<$5.8 & 
$>$1.0 & $>$6.1 &  $>0.007$ \\
\hline
\end{tabular}
\end{table}

\clearpage


\clearpage

\clearpage

\clearpage


\clearpage


\clearpage


\clearpage


\clearpage


\clearpage

\clearpage


\clearpage


\clearpage



\begin{thebibliography}{}

\bibitem[Arav(2004)]{2004ASPC..311..213A} Arav, N.\ 2004, ASP 
Conf.~Ser.~311: AGN Physics with the Sloan Digital Sky Survey, 311,
213 

\bibitem[Arav et al.(2007)]{2007ApJ...658..829A} Arav, N., et al.\ 2007, 
\apj, 658, 829 


\bibitem[AV(2009)]{} Andrade-Velazquez et al. 2009, ApJ submitted 

\bibitem{1987ApJ...320..537B}Barvainis R., 1987 ApJ 
320, 537

\bibitem{2003ApJ...598..232B} Behar, E., Rasmussen,
A.~P., Blustin, A.~J., Sako, M., Kahn, S.~M., Kaastra, J.~S.,
Branduardi-Raymont, G., \& Steenbrugge, K.~C.\ 2003, \apj, 598, 232

\bibitem[Bentz et al.(2007)]{2007ApJ...662..205B} Bentz, M.~C., et al.\ 
2007, \apj, 662, 205 

\bibitem[Blustin et 
al.(2003)]{2003A&A...403..481B} Blustin, A.~J., et al.\ 2003, \aap, 403, 481 

\bibitem[Blustin et 
al.(2005)]{2005A&A...431..111B} Blustin, A.~J., Page, M.~J., Fuerst, S.~V., Branduardi-Raymont, G., \& Ashton, C.~E.\ 2005, \aap, 431, 111 


\bibitem[Bottorff et al.(2002)]{2002ApJ...581..932B} Bottorff, M.~C., 
Baldwin, J.~A., Ferland, G.~J., Ferguson, J.~W., 
\& Korista, K.~T.\ 2002, \apj, 581, 932 


\bibitem[Chang et al.(2007)]{2007ApJ...662...94C} Chang, P., Quataert, E., 
\& Murray, N.\ 2007, \apj, 662, 94 

\bibitem[Chelouche(2008)]{2008arXiv0812.3621C} Chelouche, D.\ 2008, 
arXiv:0812.3621 


\bibitem{2003ARA&A..41..117C} Crenshaw, D.~M., 
Kraemer, S.~B., \& George, I.~M.\ 2003, \aapr, 41, 117 

\bibitem[Crenshaw et al.(2009)]{2009arXiv0902.2310C} Crenshaw, D.~M., 
Kraemer, S.~B., Schmitt, H.~R., Kaastra, J.~S., Arav, N., Gabel, J.~R., 
\& Korista, K.~T.\ 2009, arXiv:0902.2310 

\bibitem[Costantini et 
al.(2007)]{2007A&A...461..121C} Costantini, E., et al.\ 2007, \aap,
  461, 121 

\bibitem[de Vaucouleurs(1991)]{1991Sci...254.1667D} de Vaucouleurs, G.\ 
1991, Science, 254, 1667 

\bibitem[Detmers et 
al.(2008)]{2008A&A...488...67D} Detmers, R.~G., Kaastra, J.~S., Costantini, E., McHardy, I.~M., \& Verbunt, F.\ 2008, \aap, 488, 67 

\bibitem[Dorodnitsyn et al.(2008)]{2008ApJ...687...97D} Dorodnitsyn, A., 
Kallman, T., \& Proga, D.\ 2008, \apj, 687, 97 

\bibitem[Elitzur(2008)]{2008NewAR..52..274E} Elitzur, M.\ 2008, New 
Astronomy Review, 52, 274 

\bibitem{Elvis 2000} Elvis, M.\ 2000, \apj, 545, 63 

\bibitem[Everett(2005)]{2005ApJ...631..689E} Everett, J.~E.\ 2005, \apj, 
631, 689 

\bibitem{2005ApJ...634..928F} Fields, D.~L., Mathur,  S., Pogge,
  R.~W., Nicastro, F., Komossa, S., \& Krongold, Y.\ 2005, \apj,  634,
  928 

\bibitem[Fields et al.(2007)]{2007ApJ...666..828F} Fields, D.~L., Mathur, 
S., Krongold, Y., Williams, R., \& Nicastro, F.\ 2007, \apj, 666, 828 

\bibitem[Freeman et al.(2001)]{2001SPIE.4477...76F} Freeman, P., Doe, S., 
\& Siemiginowska, A.\ 2001, \procspie, 4477, 76 

\bibitem[Fruscione(2002)]{2002ChNew...9...20F} Fruscione, A.\ 2002, {\em Chandra}
News, 9, 20

\bibitem[Gabel et al.(2005)]{2005ApJ...631..741G} Gabel, J.~R., et al.\ 
2005, \apj, 631, 741 

\bibitem{19993Grev} Grevesse, M. N.,  Noels A., \& Sauval,
  A.~J. \ 1993, \aap, 271, 587

\bibitem[Hamann 
\& Ferland(1999)]{1999ARA&A..37..487H} Hamann, F., \& Ferland, G.\ 1999, \araa, 37, 487 


\bibitem[Hopkins et al.(2005)]{2005ApJ...630..705H} Hopkins, P.~F., 
Hernquist, L., Cox, T.~J., Di Matteo, T., Martini, P., Robertson, B., 
\& Springel, V.\ 2005, \apj, 630, 705 


\bibitem[Hopkins 
\& Elvis(2009)]{2009arXiv0904.0649H} Hopkins, P.~F., \& Elvis, M.\ 2009, arXiv:0904.0649 


\bibitem[Kaastra et 
al.(2004)]{2004A&A...428...57K} Kaastra, J.~S., et al.\ 2004, \aap, 428, 57 

\bibitem[Kinkhabwala et al.(2002)]{2002ApJ...575..732K} Kinkhabwala, A., et
al.\ 2002, \apj, 575, 732

\bibitem{2003ApJ...596L..27K} King, A.\ 2003, \apjl, 596, L27 

\bibitem[Komossa 
\& Breitschwerdt(2000)]{2000Ap&SS.272..299K} Komossa, S., \& Breitschwerdt, D.\ 2000, \apss, 272, 299 


\bibitem[Koyama et al.(2007)]{2007PASJ...59S..23K} Koyama, K., et al.\ 
2007, \pasj, 59, 23 

\bibitem{krol01} Krolik, J.~H.~\& Kriss, G.~A.\ 2001, \apj, 561, 684

\bibitem{kro03} Krongold, Y., Nicastro, F., Brickhouse, N.S., Elvis,
  M., Liedahl D.A. \& Mathur, S. \ 2003, \apj, 597, 832 (K03)

\bibitem{2005ApJ...620..165K} Krongold, Y.,  Nicastro, F., Elvis, M.,
  Brickhouse, N.~S., Mathur, S., \& Zezas, A.\ 2005a,  \apj, 620, 165


\bibitem{2005ApJ...622..842K} Krongold, Y.,  Nicastro, F., Brickhouse,
  N.~S., Elvis, M., \& Mathur, S.\ 2005b, \apj, 622,  842 

\bibitem[Krongold et al.(2007)]{2007ApJ...659.1022K} Krongold, Y., 
Nicastro, F., Elvis, M., Brickhouse, N., Binette, L., Mathur, S., \& 
Jim{\'e}nez-Bail{\'o}n, E.\ 2007, \apj, 659, 1022 

\bibitem[Krongold et al.(2009)]{2009ApJ...690..773K} Krongold, Y., et al.\ 
2009, \apj, 690, 773 

\bibitem[Lee et al.(2001)]{2001ApJ...554L..13L} Lee, J.~C., Ogle, P.~M., 
Canizares, C.~R., Marshall, H.~L., Schulz, N.~S., Morales, R., Fabian, 
A.~C., \& Iwasawa, K.\ 2001, \apjl, 554, L13 


\bibitem[Mathur et al.(1995)]{1995ApJ...452..230M} Mathur, S., Elvis, M., 
\& Wilkes, B.\ 1995, \apj, 452, 230 

\bibitem[Mitsuda et al.(2007)]{2007PASJ...59S...1M} Mitsuda, K., et al.\ 
2007, \pasj, 59, 1 

\bibitem[Murphy et al.(1996)]{1996ApJS..105..369M} Murphy, E.~M., Lockman, 
F.~J., Laor, A., \& Elvis, M.\ 1996, \apjs, 105, 369 

\bibitem[Nenkova et al.(2008)]{2008ApJ...685..147N} Nenkova, M., Sirocky, 
M.~M., Ivezi{\'c}, {\v Z}., \& Elitzur, M.\ 2008, \apj, 685, 147 

\bibitem[Netzer(1996)]{1996ApJ...473..781N} Netzer, H.\ 1996, \apj, 473, 
781 

\bibitem[Netzer et al.(2003)]{2003ApJ...599..933N} Netzer, H., et al.\
2003, \apj, 599, 933

\bibitem[Nicastro et al.(1999)]{1999ApJ...512..184N} Nicastro, F., Fiore, 
F., Perola, G.~C., \& Elvis, M.\ 1999, \apj, 512, 184 

\bibitem[Nicastro et al.(2000)]{2000ApJ...536..718N} Nicastro, F., et al.\ 
2000, \apj, 536, 718 


\bibitem{2004ApJ...606..151O} Ogle, P.~M., Mason, K.~O., 
Page, M.~J., Salvi, N.~J., Cordova, F.~A., McHardy, I.~M., \& Priedhorsky, 
W.~C.\ 2004, \apj, 606, 151 

\bibitem{2004ApJ...613..682P} Peterson B.M. et
  al. 2004, ApJ 613, 682

\bibitem{2005A&A...432...15P} Piconcelli, E., 
Jimenez-Bail{\'o}n, E., Guainazzi, M., Schartel, N., 
Rodr{\'{\i}}guez-Pascual, P.~M., \& Santos-Lle{\'o}, M.\ 2005, \aap, 432, 
15 

\bibitem{2004ApJ...616..688P} Proga, D., \& 
Kallman, T.~R.\ 2004, \apj, 616, 688

\bibitem[Reeves et al.(2004)]{2004ApJ...602..648R} Reeves, J.~N., Nandra, 
K., George, I.~M., Pounds, K.~A., Turner, T.~J., \& Yaqoob, T.\ 2004, \apj, 
602, 648 

\bibitem[R{\'o}{\.z}a{\'n}ska et 
al.(2008)]{2008A&A...487..895R} R{\'o}{\.z}a{\'n}ska, A., Kowalska, I., \& Gon{\c c}alves, A.~C.\ 2008, \aap, 487, 895 

\bibitem[Sako et al.(2000)]{2000ApJ...543L.115S} Sako, M., Kahn, S.~M., 
Paerels, F., \& Liedahl, D.~A.\ 2000, \apjl, 543, L115 


\bibitem{2004ApJ...608...62S} Scannapieco, E., \& 
Oh, S.~P.\ 2004, \apj, 608, 62 

\bibitem{1982ApJS...48...95S} Shull, J.~M.~\&
van Steenberg, M.\ 1982, \apjs, 48, 95

\bibitem[Steenbrugge et al.(2005)]{2005A&A...434..569S} Steenbrugge, K.~C., 
et al.\ 2005, \aap, 434, 569 

\bibitem[Suganuma et al.(2007)]{2007ASPC..373..462S} Suganuma, M., et al.\ 
2007, The Central Engine of Active Galactic Nuclei, 373, 462 

\bibitem[Turner et 
al.(2008)]{2008A&A...483..161T} Turner, T.~J., Reeves, J.~N., Kraemer, S.~B., \& Miller, L.\ 2008, \aap, 483, 161 

\bibitem{2005ARA&A..43..769V} Veilleux, S., Cecil, 
G., \& Bland-Hawthorn, J.\ 2005, \aapr, 43, 769 

\end{thebibliography}
\end{document}